\documentclass[reprint, amsmath, amssymb, aps, nofootinbib, prl]{revtex4-2}

\usepackage{graphicx}							
\usepackage{dcolumn}							
\usepackage{bm}									
\usepackage{physics}							
\usepackage{amsfonts}							
\usepackage{amssymb}							
\usepackage{tensor}								
\usepackage{dsfont}								
\usepackage{xfrac}								
\usepackage{lipsum}				
\usepackage{xcolor}
\usepackage{tcolorbox}
\usepackage[colorlinks, citecolor = magenta, linkcolor = red]{hyperref}
\usepackage{lineno}   
\usepackage{CJKutf8}  
\usepackage{mathtools}
\usepackage{svg}											

\makeatletter 
\renewcommand\onecolumngrid{
	\do@columngrid{one}{\@ne}%
	\def\set@footnotewidth{\onecolumngrid}
	\def\footnoterule{\kern-6pt\hrule width 1.5in\kern6pt}%
}

\definecolor{bluee}{HTML}{6eafdd}
\definecolor{greeen}{HTML}{009D47}

\newcommand{\jx}{\hat{J}_x}
\newcommand{\jy}{\hat{J}_y}
\newcommand{\jz}{\hat{J}_z}

\newcommand{\zh}[1]{\begin{CJK}{UTF8}{gbsn}#1\end{CJK}}

\begin{document}
	
\title{Simulating Feedback Cooling of Incoherent Quantum Mixtures}
	
\author{Kaiwen Zhu (\zh{朱凯文})}
\thanks{kaiwen.zhu@anu.edu.au}
\author{Zain Mehdi}%
\author{Joseph J. Hope}
\author{Simon A. Haine}
\affiliation{Department of Quantum Science, Research School of Physics and Engineering, \\ The Australian National University, ACT 2600, Australia}
	
\date{\today} 

\begin{abstract}
We develop a new approach for efficient and scalable simulations of measurement and control of quantum systems built upon existing phase-space methods, namely the Truncated Wigner Approximation (TWA).  We benchmark against existing particle-filter methods by simulating measurement based feedback cooling in a two-mode system, whose low-dimensional nature permits a computation of an exact solution.  The advantage of our method is multi-mode scalability, which we demonstrate through the first successful simulation of measurement-based feedback cooling of an incoherent quasi-1D thermal ensemble to quantum degeneracy. As the underlying principle of our approach exploits a general correspondence between measurement and coherent feedback, we anticipate it is also applicable across a broad range of other quantum control scenarios.
\end{abstract}
\maketitle
\textit{Introduction}\textemdash Quantum control protocols deal with the manipulation of quantum systems in order to engineer desired quantum states or to enact real-time dynamical control of the system  \cite{zhangQuantumFeedbackTheory2017, jacobsFeedbackControlNonlinear2007}.  Such methods have been employed in proposals for quantum-enhanced metrology schemes \cite{pezzeQuantumMetrologyNonclassical2018, szigetiImprovingColdatomSensors2021a}, quantum computing \cite{wisemanOptimalityFeedbackControl2008, vepsalainenImprovingQubitCoherence2022}, as well as studying dynamics in circuit and cavity QED \cite{reinerQuantumFeedbackWeakly2004, cuiFeedbackControlRabi2013}.  One example is \textit{measurement-based feedback control} (MF), where measurements performed on an ancillary system entangled with the target system are used to alter the target state \cite{wisemanQuantumTheoryFieldquadrature1993, jacobsQuantumMeasurementTheory2014, jacobsFeedbackControlNonlinear2007}.  Experimentally,  MF schemes have been employed to realise deterministic spin squeezing and noise suppression in collective spin systems \cite{coxDeterministicSqueezedStates2016, thomsenSpinSqueezingQuantum2002, inoueUnconditionalQuantumNoiseSuppression2013, hostenMeasurementNoise1002016}, feedback cooling of single atoms and micro-mechanical resonators \cite{guoFeedbackCoolingRoom2019,schafermeierQuantumEnhancedFeedback2016}, and more recently, in preparing ensembles of ultracold atoms below the shot-noise limit \cite{gajdaczPreparationUltracoldAtom2016}.

There has been particular interest in extending dynamical feedback cooling protocols to large-scale \textit{incoherent} systems; namely, feedback cooling thermal atomic gases to quantum degeneracy in order to produce Bose-Einstein Condensates (BECs) or Degenerate Fermi Gases (DFGs) \cite{hainefeedback, szigetiphasecontrast, hushControllingSpontaneousemissionNoise2013, gohFeedbackCoolingBose2022, hurstFeedbackInducedMagnetic2020, yamaguchiFeedbackcooledBoseEinsteinCondensation2023, hurstFeedbackInducedMagnetic2020, youngFeedbackstabilizedDynamicalSteady2021, munozariasSimulatingNonlinearDynamics2020a, mehdiFundamentalLimitsFeedback2024}.  
While the fundamental description of such a system is well established, simulations that simultaneously account for the multi-mode atomic dynamics, quantum correlations beyond mean-field effects and the backaction induced by the measurement process have proven elusive. This is due to the inherently large Hilbert space and difficulties of capturing finite-temperature effects associated with thermal fluctuations.

In this Letter, we introduce a highly scalable field-theoretic approach based upon existing phase-space methods \cite{psmethods, gardiner}, which is well suited for exploring the \textit{unconditional} dynamics of a wide range of controlled quantum systems.  We apply our method to investigate feedback cooling in two systems:  Firstly, a two-mode collective spin ensemble similar to those investigated in \cite{behboodFeedbackCoolingAtomic2013, munozariasSimulatingNonlinearDynamics2020a, odelliTwistandturnDynamicsSpin2024}, and secondly, cooling a quasi-1D thermal Bose gas to quantum degeneracy.  The low-dimensionality of the former permits an exact solution via Kraus operators \cite{wisemanQuantumMeasurementControl2009} against which we validate our solution, while the latter demonstrates the scalability of our approach.

Quantum systems may be mapped to phase space distributions \cite{rundleOverviewPhaseSpace2021, leeTheoryApplicationQuantum1995}, which under certain conditions can have their evolution mapped into a set of stochastic differential equations (SDEs) \cite{gardinerHandbookStochasticMethods1985, gardiner}. At the cost of requiring multiple trajectories and introducing sampling error, this approach scales like mean-field simulations while retaining the ability to include many-body correlations. This enables simulations of correlated quantum systems with non-trivial spatial structure. In degenerate quantum gases, a common choice is the Wigner representation and corresponding Truncated Wigner Approximation (TWA) \cite{wignerQuantumCorrectionThermodynamic1932, drummondTruncatedWignerDynamics2017a}.  The TWA has been shown to effectively produce a first-order correction to mean-field dynamics \cite{polkovnikovPhaseSpaceRepresentation2010}, and has been extensively used to model quantum noise \cite{steelDynamicalQuantumNoise1998, sinatraTruncatedWignerMethod2002, drummondTruncatedWignerDynamics2017, haineQuantumNoiseBright2018}, spontaneous scattering \cite{norrieQuantumTurbulenceCorrelations2006, haineSurpassingStandardQuantum2011}, optical \cite{drummondSimulationQuantumEffects1993}, atom-atom \cite{haineDynamicSchemeGenerating2009, opanchukQuantumNoiseThreedimensional2012, slodickaAtomAtomEntanglementSinglePhoton2013, lewis-swanSensitivityThermalNoise2013, haineSelfinducedSpatialDynamics2014, nolanQuantumEnhancedMeasurement2016, szigetiHighPrecisionQuantumEnhancedGravimetry2020}, and atom-light \cite{haineInformationRecyclingBeamSplitters2013, haineGenerationAtomlightEntanglement2016, kritsotakisSpinSqueezingBoseEinstein2021, fudererHybridMethodGenerating2023a} entanglement, and thermal fluctuations in ultra-cold atomic gases \cite{psmethods, blakieProjectedGrossPitaevskiiEquation2005, weilerSpontaneousVorticesFormation2008, rooneyStochasticProjectedGrossPitaevskii2012}. Quantum systems undergoing MF have conditional evolution that is correlated with the stochastic measurement result \cite{wisemanQuantumMeasurementControl2009, jacobsQuantumMeasurementTheory2014}, which can also be modelled using phase-space methods \cite{hushEfficientSimulationControlled2012}, but the measurement induces weights that reduce the relevance of trajectories that are not consistent with the measurement result. This can lead to significant sampling issues. Our approach addresses this problem by using phase-space methods to simulate the \textit{unconditional} dynamics of an equivalent, higher-dimensional quantum system.

\textit{Measurement and Coherent Feedback Schemes}\textemdash A general quantum feedback control scheme consists of the system to be controlled (e.g atomic ensemble) coupled to an ancillary system (e.g light). In MF, projective measurements are made on the ancillary system, and the measurement result $y$ is subsequently input to a controller which alters the target system through a Hamiltonian term $\hat{H}_{\text{fb}} = \sum_j f_j(y)\hat{\mathcal{O}}^{\text{s}}_{j}$ for some set of system operators $\{\hat{\mathcal{O}}_j^{\text{s}}\}$. In Coherent Feedback (CF), the explicit measurement step is absent, and control of the target system is mediated entirely through coherent interactions between the target and ancillary, achieved by engineering $\hat{H}_\mathrm{fb}$ such that the system-ancilla correlations affect the system in the desired way \cite{zhangQuantumFeedbackTheory2017, jacobsCoherentFeedbackThat2014, lloydCoherentQuantumFeedback2000}. Importantly, it has been shown that for any given MF scheme, we may always find an equivalent CF scheme which generates the same dynamics for the controlled system \cite{jacobsCoherentFeedbackThat2014}. This is true only for unconditional dynamics, as the concept of a conditional state does not exist in CF, but this lack of conditioning also means that phase-space simulations of the CF system are free of the sampling issues that constrain conditional simulations. This is precisely our approach: for a quantum system undergoing MF, we use phase-space methods to model an equivalent CF scheme, recovering the unconditional dynamics \textit{in the presence of feedback}. 

We can obtain an equivalent CF scheme by \textit{retaining} and coherently coupling the ancillary state to the system through a Hamiltonian $\hat{H}_{\text{fb}}\rightarrow \sum_j f_j(\hat{y}) \otimes\hat{\mathcal{O}}^{\text{s}}_{j}$, where $\hat{y}$ is the operator of the measured ancilla observable. Phase-space equations for the \textit{combined} system may now obtained by applying standard operator correspondences, and the result is a set of equations for \textit{both} system and ancilla phase-space variables, where feedback is implemented on each trajectory individually. This provides a stochastic simulation of the feedback-controlled system that is valid in all regimes where the phase-space simulations are valid, and can efficiently scale to systems with many modes.
\begin{figure*}[t!!]
	\centering
	\includegraphics[width = 1\textwidth]{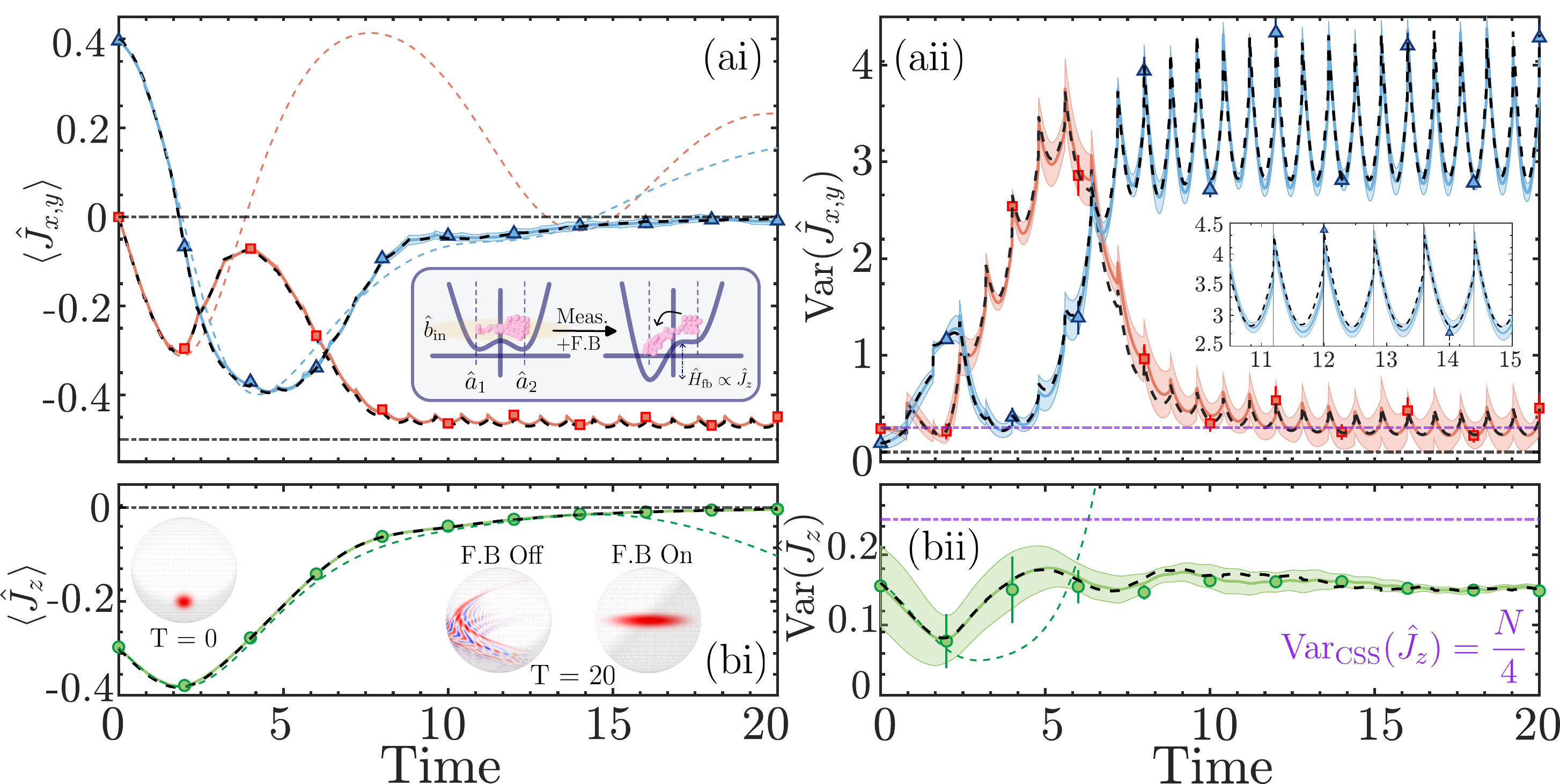}
	\caption[]{\label{paper_fig_1}Unconditional dynamics of the first two moments of the pseudo-spin observables for a CSS subject to MF damping. The initial coherent amplitudes correspond to the pseudo-spin triplet $\qty[0, 0.4, -0.3]$. All simulations were performed with parameters $\chi = 0.01, \kappa = 0.09$, $\lambda = 0.8\times 10^{-4}$, $k_{\text{fb}} = 0.1$, $N = 100$, and $\beta_0 = \sqrt{10^7}$.  A total of 62 measurements are made over the integration period.   Our method, the exact solution, and NPW are plotted using solid coloured lines, dashed black lines, and markers, respectively. Moments corresponding to $\jx$, $\jy$, and $\jz$ are plotted using red squares, blue triangles, and green circles, respectively.   Shaded regions are bootstrapped 2$\sigma$-confidence intervals in our method. (ai), (bi) Pseudo-spin means.  The dashed lines are dynamics in the absence of MF, and the SU(2) Wigner function at different times is also displayed in (bi).  The feedback potential could be realised by varying the depth of each potential well as illustrated in the inset of (ai). (aii), (bii) Pseudo-spin variances.  For reference, the purple line denotes the variance of a CSS on the equator of the Bloch sphere.} 
\end{figure*}

\textit{Feedback Scheme for a Collective Spin Ensemble}\textemdash We demonstrate and validate our approach using a two-mode atomic ensemble described by the double-well Hamiltonian $\hat{H}_\text{s} = \hbar \chi((\hat{a}_1^{\dagger}\hat{a}_1)^2 + (\hat{a}_2^{\dagger}\hat{a}_2)^2)/2 + \hbar\kappa(\hat{a}^{\dagger}_1\hat{a}_2 + \hat{a}_2^{\dagger}\hat{a}_1)$, where $\hat{a}_j$ are the annihilation operators corresponding to each mode (well) for $j\in\{1, 2\}$, and $\chi$ and $\kappa$ are the nonlinear interaction and tunneling constants, respectively.  This system realises a collective spin ensemble within the SU(2) subspace.  Double-well potentials provide a platform for studying rich many-body phenomena such as macroscopic tunneling in the bosonic Josephson effect \cite{gatiBosonicJosephsonJunction2007, milburnQuantumDynamicsAtomic1997, radzihovskyRelationAcJosephson2010}, and more recently the effectiveness of machine-learning MF through stochastic state estimation \cite{borahMeasurementBasedFeedbackQuantum2021, borahNoCollapseAccurateQuantum2023}.  The aim of the feedback is to damp out oscillations in the population difference between the wells (represented by the pseudo-spin observable $\jz = (\hat{a}^{\dagger}_1\hat{a}_1 -\hat{a}^{\dagger}_2\hat{a}_2)/2$).  This may be achieved by performing a measurement to estimate this quantity and subsequently applying the Hamiltonian $\hat{H}_{\text{fb}} = \hbar u(t)\jz$, where $u(t)$ is the feedback signal $u(t) = k_{\text{fb}}\partial_t \mathcal{J}_z^{\text{est}}$ conditioned on the time-derivative of the estimate ($\mathcal{J}_z^{\text{est}}$) with gain parameter $k_{\text{fb}}$.  We model stroboscopic Homodyne measurements separated by interval $\tau$, where the atoms are entangled with light pulses via $\hat{H}_{\text{ent}}(t) = \hbar \lambda\sum_j \Pi^{\text{ent}}_j(t, t_p)\jz \hat{b}_j^{\dagger}\hat{b}_j$ where $\lambda$ is an atom-light coupling constant, $\hat{b}_j$ is the annihilation operator corresponding to the $j$'th pulse of light and obeys $[\hat{b}_j, \hat{b}^\dagger_k] = \delta_{jk}$, and $\Pi^{\text{ent}}_j(t, t_p) = 1/t_p$ for $ t \in [t_j, t_j + t_p)$ is the unity normalised temporal mode of each pulse beginning at $t_j = j\tau$ with width $t_p \ll \tau$.  This Hamiltonian generates correlations between the atomic number difference observable with the phase quadrature of each light pulse defined as $\hat{Y}_j = i(\hat{b}_j - \hat{b}^{\dagger}_j)$.  A projective measurement of $\hat{Y}_j$ yields a result $y_j$ which is used to construct the estimate $\mathcal{J}^{\text{est}}_z(t_j) = \epsilon y_j$, where $\epsilon$ is a scale factor chosen to minimise $\expval*{(\jz(t_j) - \mathcal{J}_z^\text{est}(t_j))^2}$; for Glauber coherent states with (real) amplitude $\beta_0$, $\epsilon = (2\lambda\beta_0)^{-1}$ and $\expval*{\hat{b}_j^{\dagger}\hat{b}_j} = \abs{\beta_0}^2$.  We approximate $\partial_t \mathcal{J}^{\text{est}}_z$ via differencng consecutive measurements, which produces the MF Hamiltonian $\hat{H}_{\text{fb}} = \hbar\epsilon k_{\text{fb}}\sum_j\Pi^{\text{fb}}_j(t, \tau)(y_j - y_{j-1})$, where $\Pi_j^{\text{fb}}(t, \tau) = 1/\tau$ for $t\in (t_j + t_p, t_{j+1} + t_p]$ is the temporal mode of the $j$'th feedback operation with width $\tau$ starting \textit{after} the $j$'th entangling pulse. 
 
To derive TW equations for the system, we first obtain the equivalent CF Hamiltonian via the replacement $y_j\rightarrow\hat{Y}_j$ in $\hat{H}_{\text{fb}}$.  Applying operator correspondences and the standard truncation procedures now produces the following set of SDEs for the Wigner phase-space variables $[\alpha_1, \alpha_2, \beta_{j}]$:
\begin{align}
	i\partial_t\alpha_m &= \qty(\frac{\chi}{2}(2{|\alpha_m}|^2-1) + l_m(\lambda\mathcal{M}(t) - u(t)))\alpha_m + \kappa\alpha_n \notag\\
	i\partial_t\beta_j &= -\lambda\Pi^{\text{ent}}_j(t, t_p)\mathcal{J}_z\beta_j + i\epsilon k_{\text{fb}}\Pi^{\text{fb}}_j(t, \tau)\mathcal{J}_z, \label{eq1}
\end{align}
where $l_m = (-1)^m $, the backaction term is $\mathcal{M}(t) = \sum_j \Pi^{\text{ent}}_j(t, t_p)(\frac{1}{2} - \abs{\beta_j}^2)$, and we define the feedback signal $u(t) = \epsilon k_{\text{fb}}\sum_j \Pi^{\text{fb}}_j(t, \tau)(y_j - y_{j-1})$ where $y_j = i(\beta_j - \beta^*_j)|_{t = t_j + t_p}$.   When $t_p$ is sufficiently short compared to the timescale set by the natural atomic dynamics, the latter remains approximately constant, and the entangling dynamics may be solved analytically as $\beta_j(t_j + t_p) = \beta_j(t_j) \exp\qty(-i\lambda \mathcal{J}_z(t_j))$ and $\alpha_m(t_j + t_p) = \alpha_m(t_j)\exp\qty(-il_m\lambda(|\beta_j(t_j)|^2-\frac{1}{2}))$.  From these solutions, it is clear that measurement backaction arises from the quantum intensity fluctuations in the light coupling into the relative phase of the atomic system. 

\textit{Feedback Damping of a CSS}\textemdash We first simulate MF damping of a spin-coherent state (CSS) \cite{grossSpinSqueezingEntanglement2012, arecchiAtomicCoherentStates1972} using three methods:  Our unconditional CF approach, an exact integration using Kraus operators, and the Number Phase Wigner (NPW) particle-filter, a leading candidate for simulations of controlled quantum
systems subject to number-like measurements \cite{hushNumberphaseWignerRepresentation2010, hushEfficientSimulationControlled2012}.  The latter two methods are conditional in nature and require an additional round of averaging over independent trajectories.  Briefly, the atomic state is conditioned following each measurement as $\hat{\rho}'_\text{} = \hat{K}(y)\hat{\rho}_\text{s}\hat{K}^{\dagger}(y)$ where $\hat{K}(y) = \text{exp}(-(y-2\beta_0\text{sin}(\lambda\jz))^2/4)/(2\pi)^{1/4}$ is the Kraus operator corresponding to the quadrature measurement result $y$.  In the NPW solution, pulses of continuous stochastic weighted simulation are spaced between deterministic evolution; we provide details of implementation for both methods in the supplementary material.

\begin{figure*}[t!]
	\centering
	\includegraphics[width = 1\textwidth]{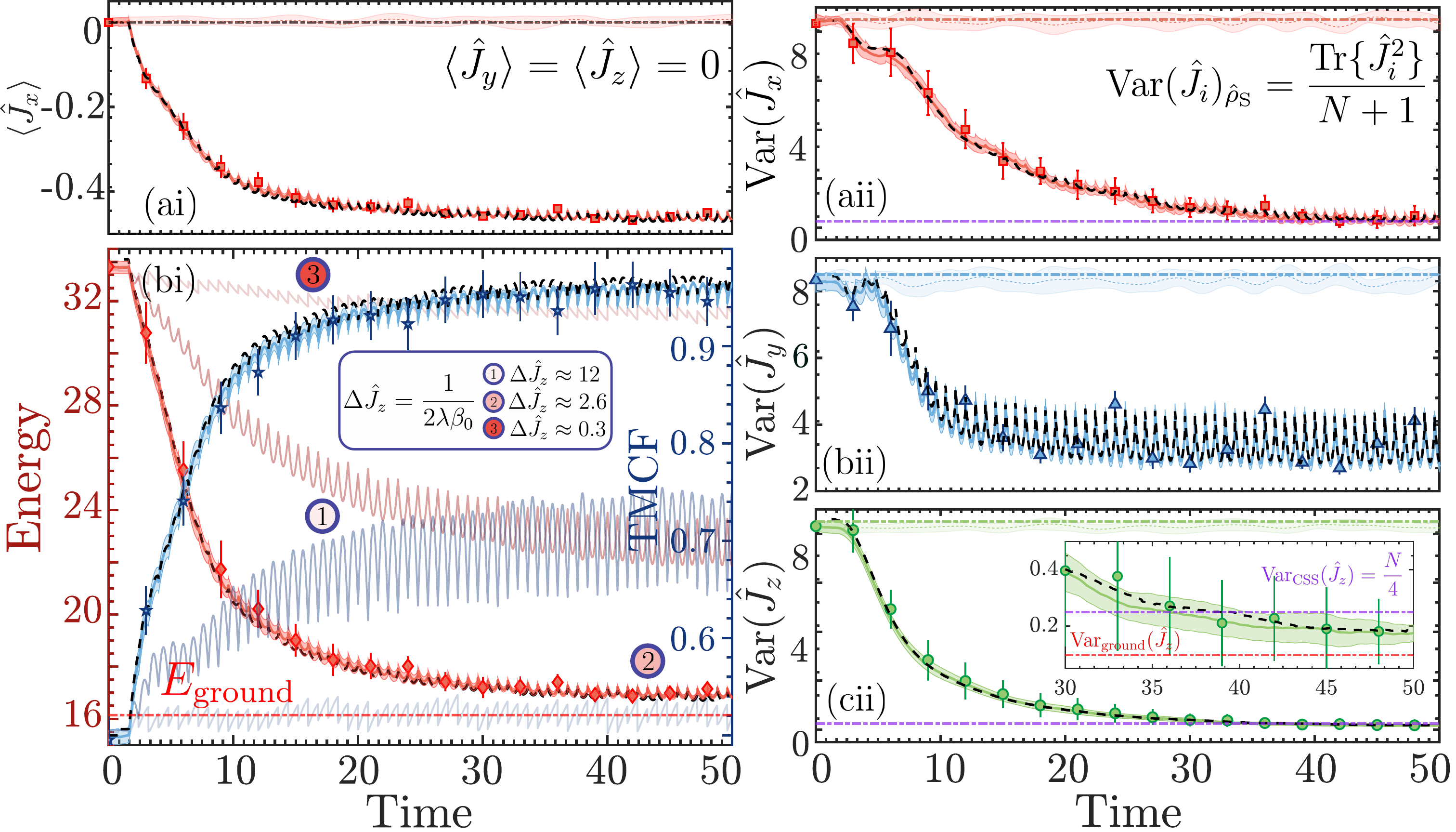}
	\caption[]{\label{paper_fig_2} (ai), (aii), (bii), (cii) Unconditional dynamics of the mean and variances of the pseudospin moments for a maximally mixed thermal spin-state subject to MF cooling.  The parameters and plotting convention are the same as those used in Fig. 1, and the dynamics in the absence of MF are also plotted.  (bi) Decrease and growth of the system energy (red) and TMCF (blue) over the MF duration, respectively, for three different measurement strengths parameterised in terms of the single-shot measurement uncertainty $\Delta \jz = 1/2\lambda\abs{\beta_0}$.}
\end{figure*}

The evolution of the first two pseudospin moments are displayed in Fig. (\ref{paper_fig_2}), where our approach exhibits quantitative agreement with both the exact solution and NPW across both orders.  We chose Hamiltonian parameters that produced non-trivial dynamics while remaining in the regime of validity for TWA where the truncation error is minimal; the dynamics in the absence of MF are plotted using dashed lines. In the inset of subfigure (aii), the variances of the conjugate observables ($\jx, \jy$) increase following each measurement due to measurement backaction, though this behaviour is absent in the variance of $\jz$ (subfigure (bii)), which approaches the steady state smoothly.  This is consistent with a quantum non-demolition measurement, where there is no backaction in the measured observable \cite{wisemanQuantumTheoryFieldquadrature1993, ilo-okekeTheorySingleShotPhase2014}.

\begin{figure}[t!]
	\centering
	\includegraphics[width = 1\columnwidth]{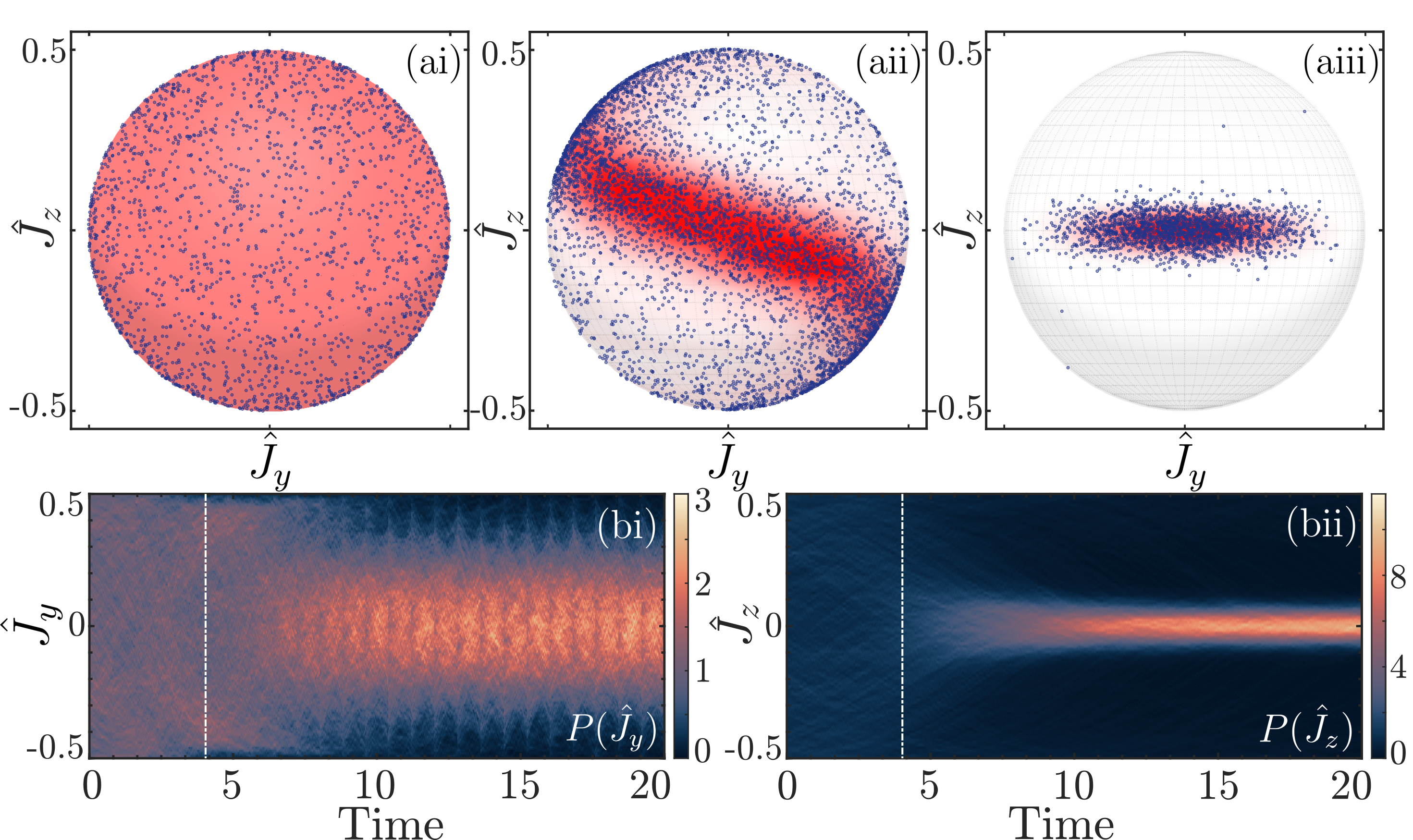}
	\caption[]{\label{paper_fig_3} Feedback cooling of a maximally mixed thermal spin state.  (ai-aiii) The exact SU(2) Wigner function and sampled CF trajectories represented on the Bloch sphere at different times (initial, intermediate, and final).  For clarity, only 2000 out of the 5000 simulated trajectories are shown.  (bi-bii) Time evolution of the $\jy$ and $\jz$ spin distributions computed from Wigner trajectories.  The dashed lines correspond to the intermediate time chosen for (aii).  The broadening of the $P(\hat{J}_y)$ distribution at measurement times due to measurement backaction is visible in (bi).}
\end{figure}

\textit{Feedback Cooling of a Spin Thermal State}\textemdash We now show that our approach is fully capable of working in incoherent regimes by simulating feedback cooling of a spin-thermal state (described by $\hat{\rho}_\text{s} = \hat{\mathbb{I}}/(N+1)$), which has been experimentally demonstrated in Ref. \cite{behboodFeedbackCoolingAtomic2013}.  The pseudospin moments are shown in Fig. (\ref{paper_fig_2}), where we again observe quantitative agreement between our approach and the exact solution.  A visual representation of the state in terms of its SU(2) Wigner function and sampled trajectories at different instances is displayed in Fig. \ref{paper_fig_3} (ai-aiii), while subfigures (bi-bii) show the evolution of the $\jy$ and $\jz$ spin distributions.

We quantify coherence by defining a two-mode condensate fraction (TMCF), borrowing from the Penrose-Onsager definition of the fraction  ($f_{\text{frac}}$) defined as the largest eigenvalue of the one-body density matrix  $\rho(x, x') = \mel*{x}{\hat{\psi}^{\dagger}(x)\hat{\psi}(x')}{x}$ normalised by the trace, such that the fraction is bounded below by $1/M$ where $M$ is the number of modes \cite{penroseBoseEinsteinCondensationLiquid1956}.  In two-modes, the matrix is $\mathcal{G}^{\text{2M}}_{nm} = \expval*{\hat{a}^{\dagger}_n\hat{a}_m}$ for $n, m \in \{1, 2\}$, so the TMCF is bounded by $\sfrac{1}{2}$. 

Evolution of the TMCF (blue) and system energy (red) is plotted in Fig. \ref{paper_fig_2} (bi) for three different measurement strengths (different transparencies) which we parameterise in terms of the single-shot measurement uncertainty $\Delta\hat{J}_z = 1/2\lambda\abs{\beta_0}$.  For the optimal strength of $\Delta\jz = 2.6$, the TMCF grows rapidly from its minimal value to a steady state value of $f^{\text{2M}}_{\text{frac}} = 0.960 \pm 0.001$, indicative of a highly pure state.  This growth is accompanied by a decrease in system energy, which approaches a steady-state energy close to that of the ground state ($E_{\text{ground}}$), and corresponds to the scenario where the opposing mechanisms of measurement backaction, and feedback, which act to increase and extract energy from the system, respectively, are balanced.  In comparison, the strengths of $\Delta\hat{J}_z = 12$ and $\Delta\hat{J}_z = 0.3$ are too weak and too strong, respectively, and both lead to sub-optimal cooling efficacy. The quality of the feedback signal is degraded in the former, while the measurement backaction (and hence heating) is too large in the latter.
\\
\begin{figure*}[t!!]
	\centering
	\includegraphics[width = 1\textwidth]{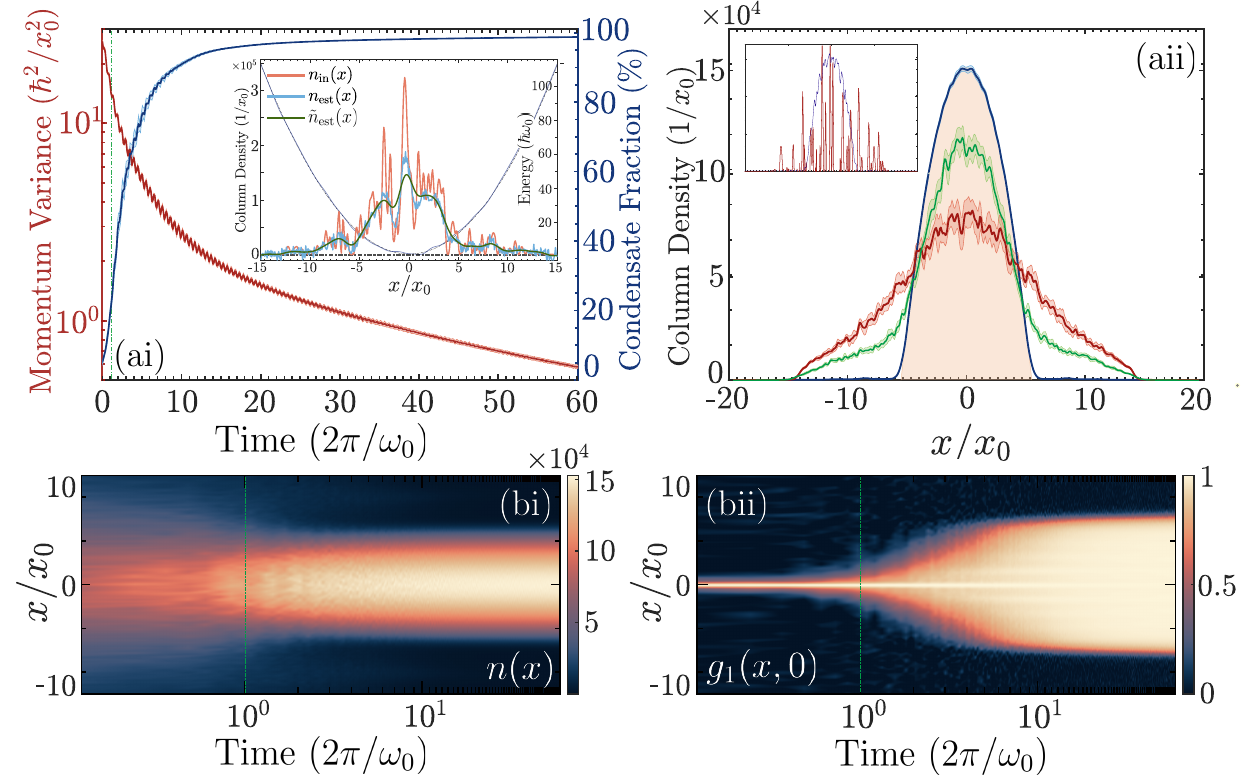}
	\caption[]{\label{paper_fig_4}(ai) Decrease of the momentum variance and growth of the condensate fraction for a quasi-1D Bose gas with $N \approx 10^6$ atoms initialised in a thermal mixture subject to MF cooling over 60 trap cycles at a rate of 150 measurements per cycle with strength $\lambda\beta_0 = 3.7\times 10^{-5}$.  The inset displays the density profiles of the real, estimate, and smoothed estimates as well as the feedback potential constructed at an intermediate time. (aii) Unconditional 1D column densities of the initial thermal state (red), an intermediate state (green) at a time corresponding to the inset in (ai), and final condensed state (blue), which agrees well with the Thomas-Fermi profile (shaded) given by $n_{\text{TF}}(x) = (\mu - V(x))/N$ where $\mu = 15.13 \ \hbar\omega_0$ is the chemical potential.  The inset shows a single Wigner trajectory for the initial (thermal) and final (cooled) state. Shaded regions represent $3\sigma$ confidence intervals. (bi-bii) Time evolution of the spatial density $(n(x))$ and first-order correlation function $(g_1(x, 0))$ across the ensemble, respectively.} 
\end{figure*}

\textit{Scalability and Multi-mode Extensions}\textemdash Finally, we demonstrate the scalability of our CF approach by simulating MF cooling of a quasi-1D incoherent thermal mixture to a BEC via non-destructive phase-contrast imaging which yield measurements of the spatial atomic density \cite{mehdiFundamentalLimitsFeedback2024, szigetiphasecontrast}.  The feedback potential is proportional to the time derivative of the spatial density estimates, and extracts energy from the system by damping out density fluctuations, and could be realised experimentally via spatio-temporal spatial-light modulators \cite{gauthierDirectImagingDigitalmicromirror2016, gauthierDynamicHighresolutionOptical2021}. 
 Previous attempts to model this process have generally employed mean-field coherent state approximations, \cite{szigetiControlledBosecondensedSources2013, yamaguchiFeedbackcooledBoseEinsteinCondensation2023} or the NPW particle-filter, which either assumes an unphysical zero-temperature coherent ensemble as the initial state, or is limited to extremely small particle numbers $(N \approx 1000)$  due to prohibitively large sampling requirements \cite{szigetiphasecontrast, gohFeedbackCoolingBose2022}.  In all cases, a quantitative, dynamical study of the coherence growth across the BEC phase transition \cite{pethickBoseEinsteinCondensation2008} which accounts for quantum correlations is lacking, which we demonstrate is possible using our approach.

Representing the multi-mode atomic state and optical field using the bosonic field operators $\hat{\psi}(x)$ and $\hat{b}(x)$, respectively, the Hamiltonian describing phase-contrast imaging for a single measurement pulse of duration $t_p$ is of the form:
\begin{align}
	\hat{H}_{\text{PC}} &= -\frac{\hbar\lambda}{t_p}\int dx \ \hat{n}(x)\int dx' \hat{b}^{\dagger}\qty(x') f^{*}\qty(x -x') \notag\\
								 & \qquad \qquad \qquad \qquad \qquad \int dx'' \ \hat{b}\qty(x'')f\qty(x-x''),
\end{align}
where $\hat{n}(x) = \hat{\psi}^{\dagger}(x)\hat{\psi}(x)$ is the atomic density operator and $f(x) = \mathcal{F}^{-1}\{\tilde{f}(k)\}$ is the diffraction-limited blurring kernel characterised in Fourier space as $f(k)\propto \text{exp} (-r_d^4\abs{k}^4/16)$, where $r_d$ is the diffraction limit of the imaging apparatus \cite{mehdiSuperfluidDissipationFeedback2024,szigetiphasecontrast, zhuFeedbackCoolingDegenerate2023, gauthierDynamicHighresolutionOptical2021}.   Applying our CF mapping, the fields (in terms of their Wigner phase-space functionals $\hat{\psi}(x)\rightarrow\psi(x)$ and $\hat{b}(x)\rightarrow\beta(x)$) transform under a single measurement as $\psi_{\text{out}}(x) = \psi_\text{in}(x)e^{-i\lambda\beta_0\mathcal{G}(x)}$ and $\beta_{\text{out}}(x) = \beta_{\text{in}}(x)e^{-i\lambda n_{\text{inf}}(x)}$, where $\mathcal{G}(x) \propto \int dx' \Theta(x')f(x-x')$ is the phase-scrambling stochastic backaction term arising from quantum fluctuations in the the optical intensity $\beta_{\text{in}}(x) = \beta_0 + \Theta(x)$, where $\Theta(x)$ is a complex Gaussian noise with correlations $\expval*{\Theta(x)\Theta^*('x)} = \delta(x-x')/2$, and $n_{\text{inf}}(x) \propto \int dx' f(x-x')n_{\text{in}}(x')$ is the diffraction-blurred density such that the quadrature signal $\mathcal{Y}(x) = i(\beta_{\text{out}}(x) - \beta^*_{\text{out}}(x)) \approx 2\lambda\beta_0 n_{\text{inf}}(x) - 2\text{Im}(\Theta(x))$ encodes a noisy estimate of the spatial density given by $n_{\text{est}}(x) = \mathcal{Y}(x)/2\lambda\beta_0$.  In practice, $n_{\text{est}}(x)$ is further convolved with a Gaussian kernel in order to capture the finite imaging and control resolution, which we denote by $\tilde{n}_{\text{est}}(x)$. In the absence of measurement, the dynamics of the atomic field are described by the cold-atom Hamiltonian \cite{pethickBoseEinsteinCondensation2008} which generates the truncated-Wigner equation of motion \cite{psmethods}:
\begin{equation}
	i\hbar \partial_t\psi(x) = (h(x) + g\abs{\psi(x)}^2)\psi(x),
\end{equation}
where $h(x) = (-\hbar^2\partial_x^2/m + m \omega_0^2 x^2)/2 + V_\text{fb}(x)$ is the single-particle Hamiltonian describing harmonic trapping with feedback potential $V_\text{fb}(x) = \hbar k_{\text{fb}}\partial_t \tilde{n}_{\text{est}}(x)$, and $g$ is the interatomic interaction strength.

In order to initialise trajectories of $\psi(x)$,  samples of a thermal state are obtained by propagating the Stochastic Projected Gross-Pitaevskii Equation (SPGPE) to equilibrium, which are the phase-space equations corresponding to a grand-canonical description accounting for energy and number-exchanging interactions between the atomic gas and a thermal reservoir \cite{rooneyStochasticProjectedGrossPitaevskii2012, rooneyNumericalMethodStochastic2014, bradleyLowdimensionalStochasticProjected2015}.  We simulate feedback cooling in 1D with stroboscopic density measurements for an ensemble of $N\approx 1.1 \times 10^6$ atoms with interaction strength $g = 0.0001 \ \hbar\omega_0 x_0$ and diffraction limit $r_d = 0.52 \ x_0$, where $x_0 = \sqrt{\hbar/m\omega_0}$ is the natural oscillator length scale. These values of $r_d$ and $g$ are consistent with the use of $^{87}$Rb atoms with trapping frequency $\omega_0 = 2\pi\times 13$ Hz.

Figure \ref{paper_fig_4} (ai) shows the growth in condensate fraction from $f_{\text{frac}} = 5.8 \pm 0.5 \%$ to $f_\text{frac} = 98.77 \pm 0.03 \%$ as well as a 70-fold reduction in the momentum variance over 60 trap periods, signaling the formation of an extremely pure condensate.  The evolution of the first order correlation function $g_1(x, 0)$ is also displayed in subfigure (bii), where it approaches unity across the spatial extent of the BEC in the steady state, which is an excellent demonstration of off-diagonal long range order \cite{pethickBoseEinsteinCondensation2008}.  Figure \ref{paper_fig_4} (aii) shows the density profiles of the initial (thermal) and final (condensed) states, the latter agreeing well with the approximate Thomas-Fermi ($T = 0 \ \text{K}$) ground state profile, as expected for a strongly interacting atomic BEC.
\\

Our findings represent the first instance where feedback cooling of atomic gases has been modeled using a full quantum-field theoretic approach starting with an incoherent thermal ensemble.  This was only possible using our new unconditional CF approach, which we have shown to be scalable to multi-mode, large-scale atomic systems where both traditional open-quantum systems methods and particle-filter methods fail. For example, a direct integration of the master equation requires a representation of the density matrix which scales as $\mathcal{O}(N^M)$ for identical bosons, which for the system represented in Fig. (\ref{paper_fig_4}) is of order $\mathcal{O}(10^{600})$. Our approach also offers a significant reduction in computational complexity; the two-mode benchmark using 500 fictitious and real NPW trajectories \cite{hushNumberphaseWignerRepresentation2010, hushEfficientSimulationControlled2012} exhausted 100 hours of CPU time, while our method using 5000 unconditional trajectories completed in five minutes, exhibiting a 1200-fold speedup.  The primary bottleneck in the former is the stochastic integrator and resampling algorithm (see appendix), which is required in all approaches derived using open-quantum systems in order to implement conditional measurement collapse and backaction in the reduced atomic system, but notably absent in our approach which instead exploits the unitary nature of CF interactions.

The natural extension of this work is to extend the MF scheme for the quasi-1D Bose gas to higher dimensional geometries; though our one-dimensional results are promising and may offer insight towards optimal parameter regimes for efficient cooling, extensive variables such as the final BEC size or condensate fraction may only be interpreted as physical thermodynamic quantities in 3D \cite{pethickBoseEinsteinCondensation2008, pitaevskiiBoseEinsteinCondensation2003}. In addition, our approach may be adapted to incorporate measurement-induced spontaneous emission losses \cite{kritsotakisSpinSqueezingBoseEinstein2021}, which is the primary limiting factor in the efficacy of feedback cooling from high-energy thermal states \cite{mehdiFundamentalLimitsFeedback2024}. Finally, due to built-in scalability and the ability to generalise our approach to model continuous measurements, we anticipate it may be used in other physical systems such as quantum optomechanics \cite{aspelmeyerCavityOptomechanics2014}, as well as across the broader field of quantum control scenarios such as reservoir engineering \cite{poyatosQuantumReservoirEngineering1996, basilewitschReservoirEngineeringUsing2019, tissotReservoirEngineeringClassical2024}, state preparation \cite{handelModellingFeedbackControl2005, millsHighFidelityStatePreparation2022}, active stabilisation \cite{sayrinRealtimeQuantumFeedback2011, guoActivefeedbackQuantumControl2023}, and also investigating large-scale collective many-body phenomena such as the quantum-classical transition in chaotic quantum systems \cite{eastmanControllingChaosQuantum2019, munozariasSimulatingNonlinearDynamics2020a}.
\\

The source code and data for all simulations presented in this paper may be found at \cite{Zhu_Feedback_Cooling_Code_quackle}.
\\

\textit{Acknowledgments}\textemdash We acknowledge the Ngunnawal people as the traditional custodians of the land upon which this research was conducted, and recognise that sovereignty was never ceeded. We thank Australian Research Council Project No. DP190101709, the National Computational Infrastructure for access to computing resources, and  Stuart Szigeti for insightful discussions.  SAH acknowledges support through an Australian Research Council Future Fellowship Grant No. FT210100809. KZ thanks the support provided by the D.N.F Dunbar Honours Scholarship and Australian Government Research Training Program. KZ would also like to thank his nana Jun Hua (\zh{华军}) for insightful discussion on the scope of this work as well as ongoing support and devotion in all other aspects of his life, too.

\newpage
\section{Appendix}
\textit{Number-Phase Wigner Implementation}\textemdash The atomic NPW equations closely resembles Eq. (\ref{eq1}) once the optical terms $(\beta_j)$ are discarded.  The Hamiltonian terms are identical, while the backaction and weights evolution are as follows:
\begin{align}\label{eq5}
	d\alpha_m^j &= \hdots - \frac{i\sqrt{\gamma}}{2}l_m\alpha^j_m \circ dV^j  \notag\\
	\frac{d\omega^j}{\omega^j} &= -2\gamma(\mathcal{J}_z - \expval*{\jz})^2 dt + 2\sqrt{\gamma}\mathcal{J}_z\circ dW,
\end{align}
where $\circ \  dV$ and $\circ \  dW$ are real Wiener noises in the Stratonovich calculus and $\expval*{f(\hat{\bm{a}})} = \mathbb{E}_c[f(\bm{\alpha}^j)] = \sum_j \omega^jf(\bm{\alpha})/\sum_j\omega^{j}$ are conditional expectation values over a swarm of fictitious trajectories indexed by $j$.  For each measurement, we evolve Eq. (\ref{eq5}) continuously over a period $t_p^{\text{NPW}}$ such that $t_{p}^{\text{NPW}}/dt_{\text{CF}} \ll 1$ which allows us to update the trajectories instantaneously over a single $dt_{\text{CF}}$ timestep.  The measurement result over a single pulse $t^{\text{NPW}}_p$ is obtained by integrating over the continuous measurement current $dy = \expval*{\jz}dt + dW/(2\sqrt{\gamma})$ in order to produce $\mathcal{J}_{z_{\text{est}}}^{\text{NPW}} = (\int_{t_0}^{t_0 + t_p^{\text{NPW}}}dy)/t_p^{\text{NPW}}$, where ($\gamma$) is the measurement rate parameter which satisfies the equation $\gamma t_p^{\text{NPW}} = \lambda^2\beta_0^2$.  This equality may be obtained by either equating the CF and NPW backaction terms (i.e Eqs. (\ref{eq1}) and (\ref{eq5})) and enforcing that the noise terms $\circ \ dV$ and $\Theta_j$ obey the same distribution, where $\Theta_j$ is a complex Gaussian noise characterised by $\expval*{\Theta(t)_j\Theta(t')^*_k} = \Theta(t-t')\delta_{jk}/2$ which appears in the initial sampling of the light variables $\beta_j = \beta_0 + \Theta_j$, or by equating the SNR of a single-shot measurement.

In practice, we evolve a swarm of 500 fictitious trajectories (each with their own noise $dV^j$) under the \textit{same} measurement noise $dW$ to simulate a single conditional trajectory.  Over the measurement period, trajectory weights decay exponentially with respect to the difference between their $\mathcal{J}_z$ values and the measurement record.  This rapidly leads to a single trajectory dominating, and so the state is severely undersampled and not well-represented.  This is referred to in the literature as \textit{particle-impoverishment} \cite{liDeterministicResamplingUnbiased2012}, and we employ the Kitigawa sequential importance resampling algorithm to address this problem \cite{kitagawaMonteCarloFilter1996}.  

\textit{Kraus Operator Implementation}\textemdash Let the system ($\hat{\rho}_{\text{s}}$) and ancilla ($\hat{\rho}_{\text{a}}$) be the atomic ensemble and light, respectively.  An arbitrary atom-light state ($\hat{\rho}_0 = \hat{\rho}_\text{s}\otimes \hat{\rho}_{\text{a}}$) expressed in the Dicke basis as $\hat{\rho}_0 = \sum_{n, m = -N/2}^{N/2}\rho_{nm}\dyad{n}{m}\otimes\dyad{\beta_0}{\beta_0}$, where $\rho_{nm}$ are coefficients for the atomic state and we have chosen the light as a coherent state with amplitude $\beta_0$.  The post-entanglement state is given by $\hat{\rho}_{\text{ent}} = \hat{U}_{\text{ent}}\hat{\rho}_0\hat{U}^{\dagger}_{\text{ent}}$ where $\hat{U}_{\text{ent}} = \text{exp}(-i\lambda \jz \otimes \hat{b}^{\dagger}\hat{b})$ and is equal to:
\begin{align}
	\hat{\rho}_{\text{ent}} = \sum_{n, m}\rho_{nm}\dyad{n}{m}\otimes\dyad*{\beta_0 e^{i\lambda n}}{\beta_0 e^{i \lambda m}}.
\end{align}
Prior to each measurement, we construct the probability distribution $P(y, t) = \text{Tr}_l(\hat{\rho}_\text{ent}(t)\dyad{y}{y})$ which takes the form:
\begin{align}
	P\qty(y, t)= \sum_{n = -N/2}^{N/2}\frac{\rho_{nn}(t)}{\sqrt{\pi}}\exp\biggl(-\frac{1}{2}\qty(y - 2\beta_0\sin\qty(\lambda n))^2\biggr),
\end{align}
from which we generate a single measurement result $y$ using rejection sampling.  To condition $\hat{\rho}_{\text{ent}}$ onto result $y$, we trace out the light and apply the Kraus operator $\hat{K}(y) = \text{exp}(-(y-2\beta_0\text{sin}(\lambda\jz))^2/4)/(2\pi)^{1/4}$ to obtain $\hat{\rho}_y = \hat{K}(y)\text{Tr}_l(\hat{\rho}_\text{ent})\hat{K}^{\dagger}(y)$ with appropriate renormalisation, which is taken as the new atomic state at the beginning of the next timestep.  Between measurements, the state is evolved as $\hat{\rho}_\text{a}(t + \Delta t) = \hat{U}(\Delta t)\hat{\rho}_\text{a}(t)\hat{U}^{\dagger}(\Delta t)$ where $\hat{U}(\Delta t) = \text{exp}(-i(\hat{H}_\text{s} + \hat{H}_{\text{fb}})\Delta t/\hbar)$.

\textit{TWA Implementation}\textemdash The backaction term ($\mathcal{M}(t)$) in Eq. (\ref{eq1}) imprinted onto the relative phase of the atomic system comprises a deterministic and stochastic rotation about the $\jz$ axis due to the coherent amplitude and quantum intensity noise.  In practice, we can always cancel out the former by applying a counter rotation with magnitude  $\lambda\abs{\beta_0}^2$ after each entangling pulse.

The SU(2) subspace is number conserving, while TW samples capture number fluctuations.  To ensure fair comparison between the CF and exact solutions, we manually normalise trajectories to $N$, i.e $|\alpha^j_1|^2 + |\alpha^j_2|^2 = N$.

\textit{Quasi-1D Thermal State Initialisation}\textemdash The complete SPGPE accounts for both energy and number exchanging processes between a coherent `\textbf{C}-field' and thermal reservoir \cite{bradleyLowdimensionalStochasticProjected2015, rooneyStochasticProjectedGrossPitaevskii2012}.  For initialising thermal trajectories, it is sufficient to neglect the energy-damping terms, and the remaining terms form the \textit{simple-growth} SPGPE:
\begin{equation}\label{eq8}
	d\psi|_{\text{SG}} = \hdots + \mathcal{P}\{\gamma (\mu - \mathcal{L})\psi dt + dW_\gamma(x, t)\},
\end{equation}
where we have defined the Liovillian operator $\mathcal{L}\psi = (\hat{H}_{\text{CA}} + g\abs{\psi}^2\psi)\psi$ and $dW_{\gamma}(x, t)$ is a complex Wiener noise characterised by the correlations $\expval*{dW^{*}_{\gamma}(x, t)dW_{\gamma}(x', t)} = 2\gamma \Tilde{T}\delta(x-x')dt$ where $\gamma$ and $\tilde{T} = k_\text{b}T/\hbar\omega_0$ are growth rate and dimensionless temperature parameters, respectively.  To obtain thermal samples, we evolve Eq. (\ref{eq8}) to equilibrium in the Hermite-Gauss basis with the projector $\mathcal{P}\{\star\}$ restricting evolution into $n = 100$ modes.  

In theory, increasing $\tilde{T}$ should lead to thermal states with near-zero condensate fraction.  However, the 1D reduction prevents the existence of `true' thermal states, and we find an effective lower bound of $f_{\text{frac}} \approx 5-6\%$ regardless of parameter choice or the number of modes $n$.  As the simple-growth SPGPE realises a grand-canonical description, we adjust the particle number ($N$) at equilibrium simply by varying parameters $\mu$ and $\tilde{T}$. 

\textit{Equivalence of MF and CF Unconditional State} \textemdash We now show that a single iteration of MF and CF lead to identical unconditional states for the system.  Generalising Eq. (\ref{eq5}),  the combined system-ancilla state post entanglement is described by $\hat{\rho}_{\text{ent}} = \hat{U}_{\text{ent}}\qty(\hat{\rho}_{\text{s}}\otimes \hat{\rho}_{\text{a}})\hat{U}^{\dagger}_{\text{ent}}$, Assuming a pure state for the ancillary (i.e $\hat{\rho}_\text{a} = \dyad{\psi_\text{a}}{\psi_{\text{a}}}$), a projective measurement of an ancillary observable ($\hat{y}$) conditions the system state (onto the measurement result defined by $\hat{y}\ket{y} = y\ket{y}$):
\begin{align}
	\hat{\rho}_{\text{s}}(y) &= \mathcal{N}(y)\text{Tr}_\text{a}(\dyad{y}{y}\hat{\rho}_{\text{ent}}\dyad{y}{y}) \notag\\
		&= \mathcal{N}(y)\mel{\psi_\text{a}}{\hat{U}_{\text{ent}}}{y} \hat{\rho}_{\text{s}} \mel{y}{\hat{U}^{\dagger}_{\text{ent}}}{\psi_{\text{a}}}.
\end{align}
We have traced out the ancillary and introduced the normalisation factor $\mathcal{N}(y) = \text{Tr}(\hat{K}^{\dagger}(y)\hat{K}(y)\hat{\rho}_\text{s})^{-1}$, where $\hat{K}(y) = \mel{\psi_\text{a}}{\hat{U}_{\text{ent}}}{y}$ is an equivalent expression for the Kraus operator.  Under MF, we apply the conditional feedback Hamiltonian $\hat{H}_{\mathrm{fb}}(y) = \sum_j f_j(y)\hat{\mathcal{O}}_j^{\text{s}}$ to $\hat{\rho}_{\text{s}}(y)$ for some time $\Delta t$ to obtain the conditional feedback state $\hat{\rho}^{\text{fb}}_{\text{s}}(y) = \hat{U}^{\text{MF}}_{\text{fb}}(y)\hat{\rho}_\text{s}(y)\hat{U}^{\text{MF}\dagger}_{\text{fb}}(y)$, where $\hat{U}^{\text{MF}}_{\text{fb}}(y) = \text{exp}(-i\hat{H}_{\text{fb}}(y)\Delta t/\hbar)$.  The \textit{unconditional} feedback state is recovered by averaging over all conditional states:
\begin{align}\label{eq9}
	\hat{\rho}_\text{s, MF}^{\text{fb}} &= \int dy \  P(y) \hat{U}^{\text{MF}}_{\text{fb}}(y)\hat{\rho}_\text{s}(y)\hat{U}^{\text{MF}\dagger}_{\text{fb}}(y),
\end{align}
where $P(y) = \mathcal{N}(y)^{-1}$ is the probability distribution of measurement results.

In CF, no projective measurement is performed on the ancillary. Instead, an \textit{unconditional} feedback Hamiltonian $\hat{H}_{\mathrm{fb}}(\hat{y}) = \sum_j f_j(\hat{y})\otimes\hat{\mathcal{O}}_j^{\text{s}}$ which acts on the \textit{combined} system-ancillary space is directly applied to the post-entanglement state ($\hat{\rho}_{\text{ent}}$) to obtain the unconditional \textit{combined} feedback state $\hat{\rho}^{\text{fb}} = \hat{U}^{\text{CF}}_{\text{fb}}\hat{U}_{\text{ent}}\qty(\hat{\rho}_{\text{s}}\otimes \hat{\rho}_{\text{a}})\hat{U}^{\dagger}_{\text{ent}}\hat{U}^{\text{CF}\dagger}_{\text{fb}}$, where $\hat{U}^{\text{CF}}_{\text{fb}} = \text{exp}(-i\Delta t\hat{H}_{\text{fb}}\Delta t/\hbar)$.  We now show that the unconditional feedback state for the \textit{system} is identical to that of the MF scheme in Eq. (\ref{eq9}).  Firstly, expanding $\hat{U}^{\text{CF}}_{\text{fb}}$ in the spectral basis of the ancillary yields $\hat{U}^{\text{CF}}_{\text{fb}} = \text{exp}(-i\Delta t\sum_j f_j(\hat{y})\otimes \hat{\mathcal{O}}^{\text{s}}_j/\hbar)
= \int dy \ \text{exp}(-i\Delta t\sum_j f_j(y) \hat{\mathcal{O}}^{\text{s}}_j/\hbar)\otimes \dyad{y}{y}
= \int dy \ \hat{U}^{\text{MF}}_{\text{fb}}(y)\otimes \dyad{y}{y}$.  Tracing over the ancillary, noting that  $\comm*{\hat{y}}{\hat{\mathcal{O}}_j^\text{s}} = 0$ we obtain:
\begin{align}
	\hat{\rho}^{\text{fb}}_{\text{s, CF}} &= \int dy_1 \ \mel{y_1}{\hat{U}^{\text{CF}}_{\text{fb}}\hat{U}_{\text{ent}}\qty(\hat{\rho}_{\text{s}}\otimes \hat{\rho}_{\text{a}})\hat{U}^{\dagger}_{\text{ent}}\hat{U}^{\text{CF}\dagger}_{\text{fb}}}{y_1}\notag \\
	&=\int d\bm{y}\braket{y_1}{y_2}\hat{U}^{\text{MF}}_{\text{fb}}(y_2)\hat{K}(y_2)\hat{\rho}_{\text{s}}\hat{K}^{\dagger}(y_3)\hat{U}^{\text{MF}\dagger}_{\text{fb}}(y_3)\braket{y_3}{y_1}\notag \\
	&= \int dy \  \mathcal{N}(y)^{-1}\hat{U}^{\text{MF}}_{\text{fb}}(y)(\mathcal{N}(y)\hat{K}(y)\hat{\rho}_{\text{s}}\hat{K}^{\dagger}(y))\hat{U}^{\text{MF}\dagger}_{\text{fb}}(y)\notag\\
	&=\int dy \ P(y) \hat{U}^{\text{MF}}_{\text{fb}}(y)\hat{\rho}_{\text{s}}(y)\hat{U}^{\text{MF}\dagger}_{\text{fb}}(y) \equiv \hat{\rho}_{\text{s, MF}}^{\text{fb}}.
\end{align}
Hence, under our mapping $\hat{H}_{\text{fb}}(y)\rightarrow\hat{H}_{\text{fb}}(\hat{y})$, the unconditional system state in MF obtained after a single (and by extension, a series of) feedback iteration(s) may be directly recovered from our proposed CF scheme. We emphasise that the equivalent CF scheme need not be experimentally realistic; as long as the Hamiltonian is Hermitian, it is \textit{in principle}, physical. 
%



\bibliography{zhu_k1_ver_arxiv}

\begin{thebibliography}{92}%
\makeatletter
\providecommand \@ifxundefined [1]{%
 \@ifx{#1\undefined}
}%
\providecommand \@ifnum [1]{%
 \ifnum #1\expandafter \@firstoftwo
 \else \expandafter \@secondoftwo
 \fi
}%
\providecommand \@ifx [1]{%
 \ifx #1\expandafter \@firstoftwo
 \else \expandafter \@secondoftwo
 \fi
}%
\providecommand \natexlab [1]{#1}%
\providecommand \enquote  [1]{``#1''}%
\providecommand \bibnamefont  [1]{#1}%
\providecommand \bibfnamefont [1]{#1}%
\providecommand \citenamefont [1]{#1}%
\providecommand \href@noop [0]{\@secondoftwo}%
\providecommand \href [0]{\begingroup \@sanitize@url \@href}%
\providecommand \@href[1]{\@@startlink{#1}\@@href}%
\providecommand \@@href[1]{\endgroup#1\@@endlink}%
\providecommand \@sanitize@url [0]{\catcode `\\12\catcode `\$12\catcode
  `\&12\catcode `\#12\catcode `\^12\catcode `\_12\catcode `\%12\relax}%
\providecommand \@@startlink[1]{}%
\providecommand \@@endlink[0]{}%
\providecommand \url  [0]{\begingroup\@sanitize@url \@url }%
\providecommand \@url [1]{\endgroup\@href {#1}{\urlprefix }}%
\providecommand \urlprefix  [0]{URL }%
\providecommand \Eprint [0]{\href }%
\providecommand \doibase [0]{https://doi.org/}%
\providecommand \selectlanguage [0]{\@gobble}%
\providecommand \bibinfo  [0]{\@secondoftwo}%
\providecommand \bibfield  [0]{\@secondoftwo}%
\providecommand \translation [1]{[#1]}%
\providecommand \BibitemOpen [0]{}%
\providecommand \bibitemStop [0]{}%
\providecommand \bibitemNoStop [0]{.\EOS\space}%
\providecommand \EOS [0]{\spacefactor3000\relax}%
\providecommand \BibitemShut  [1]{\csname bibitem#1\endcsname}%
\let\auto@bib@innerbib\@empty
\bibitem [{\citenamefont {Zhang}\ \emph {et~al.}(2017)\citenamefont {Zhang},
  \citenamefont {Liu}, \citenamefont {Wu}, \citenamefont {Jacobs},\ and\
  \citenamefont {Nori}}]{zhangQuantumFeedbackTheory2017}%
  \BibitemOpen
  \bibfield  {author} {\bibinfo {author} {\bibfnamefont {J.}~\bibnamefont
  {Zhang}}, \bibinfo {author} {\bibfnamefont {Y.-x.}\ \bibnamefont {Liu}},
  \bibinfo {author} {\bibfnamefont {R.-B.}\ \bibnamefont {Wu}}, \bibinfo
  {author} {\bibfnamefont {K.}~\bibnamefont {Jacobs}},\ and\ \bibinfo {author}
  {\bibfnamefont {F.}~\bibnamefont {Nori}},\ }\bibfield  {title} {\bibinfo
  {title} {Quantum feedback: {{Theory}}, experiments, and applications},\
  }\href {https://doi.org/10.1016/j.physrep.2017.02.003} {\bibfield  {journal}
  {\bibinfo  {journal} {Physics Reports}\ }\bibinfo {series} {Quantum Feedback:
  Theory, Experiments, and Applications},\ \textbf {\bibinfo {volume} {679}},\
  \bibinfo {pages} {1} (\bibinfo {year} {2017})}\BibitemShut {NoStop}%
\bibitem [{\citenamefont {Jacobs}\ and\ \citenamefont
  {Lund}(2007)}]{jacobsFeedbackControlNonlinear2007}%
  \BibitemOpen
  \bibfield  {author} {\bibinfo {author} {\bibfnamefont {K.}~\bibnamefont
  {Jacobs}}\ and\ \bibinfo {author} {\bibfnamefont {A.~P.}\ \bibnamefont
  {Lund}},\ }\bibfield  {title} {\bibinfo {title} {Feedback {{Control}} of
  {{Nonlinear Quantum Systems}}: {{A Rule}} of {{Thumb}}},\ }\href
  {https://doi.org/10.1103/PhysRevLett.99.020501} {\bibfield  {journal}
  {\bibinfo  {journal} {Physical Review Letters}\ }\textbf {\bibinfo {volume}
  {99}},\ \bibinfo {pages} {020501} (\bibinfo {year} {2007})}\BibitemShut
  {NoStop}%
\bibitem [{\citenamefont {Pezz{\`e}}\ \emph {et~al.}(2018)\citenamefont
  {Pezz{\`e}}, \citenamefont {Smerzi}, \citenamefont {Oberthaler},
  \citenamefont {Schmied},\ and\ \citenamefont
  {Treutlein}}]{pezzeQuantumMetrologyNonclassical2018}%
  \BibitemOpen
  \bibfield  {author} {\bibinfo {author} {\bibfnamefont {L.}~\bibnamefont
  {Pezz{\`e}}}, \bibinfo {author} {\bibfnamefont {A.}~\bibnamefont {Smerzi}},
  \bibinfo {author} {\bibfnamefont {M.~K.}\ \bibnamefont {Oberthaler}},
  \bibinfo {author} {\bibfnamefont {R.}~\bibnamefont {Schmied}},\ and\ \bibinfo
  {author} {\bibfnamefont {P.}~\bibnamefont {Treutlein}},\ }\bibfield  {title}
  {\bibinfo {title} {Quantum metrology with nonclassical states of atomic
  ensembles},\ }\href {https://doi.org/10.1103/RevModPhys.90.035005} {\bibfield
   {journal} {\bibinfo  {journal} {Reviews of Modern Physics}\ }\textbf
  {\bibinfo {volume} {90}},\ \bibinfo {pages} {035005} (\bibinfo {year}
  {2018})},\ \Eprint {https://arxiv.org/abs/1609.01609} {arXiv:1609.01609
  [quant-ph]} \BibitemShut {NoStop}%
\bibitem [{\citenamefont {Szigeti}\ \emph {et~al.}(2021)\citenamefont
  {Szigeti}, \citenamefont {Hosten},\ and\ \citenamefont
  {Haine}}]{szigetiImprovingColdatomSensors2021a}%
  \BibitemOpen
  \bibfield  {author} {\bibinfo {author} {\bibfnamefont {S.~S.}\ \bibnamefont
  {Szigeti}}, \bibinfo {author} {\bibfnamefont {O.}~\bibnamefont {Hosten}},\
  and\ \bibinfo {author} {\bibfnamefont {S.~A.}\ \bibnamefont {Haine}},\
  }\bibfield  {title} {\bibinfo {title} {Improving cold-atom sensors with
  quantum entanglement: {{Prospects}} and challenges},\ }\href
  {https://doi.org/10.1063/5.0050235} {\bibfield  {journal} {\bibinfo
  {journal} {Applied Physics Letters}\ }\textbf {\bibinfo {volume} {118}},\
  \bibinfo {pages} {140501} (\bibinfo {year} {2021})}\BibitemShut {NoStop}%
\bibitem [{\citenamefont {Wiseman}\ and\ \citenamefont
  {Bouten}(2008)}]{wisemanOptimalityFeedbackControl2008}%
  \BibitemOpen
  \bibfield  {author} {\bibinfo {author} {\bibfnamefont {H.~M.}\ \bibnamefont
  {Wiseman}}\ and\ \bibinfo {author} {\bibfnamefont {L.}~\bibnamefont
  {Bouten}},\ }\bibfield  {title} {\bibinfo {title} {Optimality of {{Feedback
  Control Strategies}} for {{Qubit Purification}}},\ }\href
  {https://doi.org/10.1007/s11128-008-0075-8} {\bibfield  {journal} {\bibinfo
  {journal} {Quantum Information Processing}\ }\textbf {\bibinfo {volume}
  {7}},\ \bibinfo {pages} {71} (\bibinfo {year} {2008})}\BibitemShut {NoStop}%
\bibitem [{\citenamefont {Veps{\"a}l{\"a}inen}\ \emph
  {et~al.}(2022)\citenamefont {Veps{\"a}l{\"a}inen}, \citenamefont {Winik},
  \citenamefont {Karamlou}, \citenamefont {Braum{\"u}ller}, \citenamefont
  {Paolo}, \citenamefont {Sung}, \citenamefont {Kannan}, \citenamefont
  {Kjaergaard}, \citenamefont {Kim}, \citenamefont {Melville}, \citenamefont
  {Niedzielski}, \citenamefont {Yoder}, \citenamefont {Gustavsson},\ and\
  \citenamefont {Oliver}}]{vepsalainenImprovingQubitCoherence2022}%
  \BibitemOpen
  \bibfield  {author} {\bibinfo {author} {\bibfnamefont {A.}~\bibnamefont
  {Veps{\"a}l{\"a}inen}}, \bibinfo {author} {\bibfnamefont {R.}~\bibnamefont
  {Winik}}, \bibinfo {author} {\bibfnamefont {A.~H.}\ \bibnamefont {Karamlou}},
  \bibinfo {author} {\bibfnamefont {J.}~\bibnamefont {Braum{\"u}ller}},
  \bibinfo {author} {\bibfnamefont {A.~D.}\ \bibnamefont {Paolo}}, \bibinfo
  {author} {\bibfnamefont {Y.}~\bibnamefont {Sung}}, \bibinfo {author}
  {\bibfnamefont {B.}~\bibnamefont {Kannan}}, \bibinfo {author} {\bibfnamefont
  {M.}~\bibnamefont {Kjaergaard}}, \bibinfo {author} {\bibfnamefont {D.~K.}\
  \bibnamefont {Kim}}, \bibinfo {author} {\bibfnamefont {A.~J.}\ \bibnamefont
  {Melville}}, \bibinfo {author} {\bibfnamefont {B.~M.}\ \bibnamefont
  {Niedzielski}}, \bibinfo {author} {\bibfnamefont {J.~L.}\ \bibnamefont
  {Yoder}}, \bibinfo {author} {\bibfnamefont {S.}~\bibnamefont {Gustavsson}},\
  and\ \bibinfo {author} {\bibfnamefont {W.~D.}\ \bibnamefont {Oliver}},\
  }\bibfield  {title} {\bibinfo {title} {Improving qubit coherence using
  closed-loop feedback},\ }\href {https://doi.org/10.1038/s41467-022-29287-4}
  {\bibfield  {journal} {\bibinfo  {journal} {Nature Communications}\ }\textbf
  {\bibinfo {volume} {13}},\ \bibinfo {pages} {1932} (\bibinfo {year}
  {2022})}\BibitemShut {NoStop}%
\bibitem [{\citenamefont {Reiner}\ \emph {et~al.}(2004)\citenamefont {Reiner},
  \citenamefont {Smith}, \citenamefont {Orozco}, \citenamefont {Wiseman},\ and\
  \citenamefont {Gambetta}}]{reinerQuantumFeedbackWeakly2004}%
  \BibitemOpen
  \bibfield  {author} {\bibinfo {author} {\bibfnamefont {J.~E.}\ \bibnamefont
  {Reiner}}, \bibinfo {author} {\bibfnamefont {W.~P.}\ \bibnamefont {Smith}},
  \bibinfo {author} {\bibfnamefont {L.~A.}\ \bibnamefont {Orozco}}, \bibinfo
  {author} {\bibfnamefont {H.~M.}\ \bibnamefont {Wiseman}},\ and\ \bibinfo
  {author} {\bibfnamefont {J.}~\bibnamefont {Gambetta}},\ }\bibfield  {title}
  {\bibinfo {title} {Quantum feedback in a weakly driven cavity {{QED}}
  system},\ }\href {https://doi.org/10.1103/PhysRevA.70.023819} {\bibfield
  {journal} {\bibinfo  {journal} {Physical Review A}\ }\textbf {\bibinfo
  {volume} {70}},\ \bibinfo {pages} {023819} (\bibinfo {year}
  {2004})}\BibitemShut {NoStop}%
\bibitem [{\citenamefont {Cui}\ and\ \citenamefont
  {Nori}(2013)}]{cuiFeedbackControlRabi2013}%
  \BibitemOpen
  \bibfield  {author} {\bibinfo {author} {\bibfnamefont {W.}~\bibnamefont
  {Cui}}\ and\ \bibinfo {author} {\bibfnamefont {F.}~\bibnamefont {Nori}},\
  }\bibfield  {title} {\bibinfo {title} {Feedback control of {{Rabi}}
  oscillations in circuit {{QED}}},\ }\href
  {https://doi.org/10.1103/PhysRevA.88.063823} {\bibfield  {journal} {\bibinfo
  {journal} {Physical Review A}\ }\textbf {\bibinfo {volume} {88}},\ \bibinfo
  {pages} {063823} (\bibinfo {year} {2013})}\BibitemShut {NoStop}%
\bibitem [{\citenamefont {Wiseman}\ and\ \citenamefont
  {Milburn}(1993)}]{wisemanQuantumTheoryFieldquadrature1993}%
  \BibitemOpen
  \bibfield  {author} {\bibinfo {author} {\bibfnamefont {H.~M.}\ \bibnamefont
  {Wiseman}}\ and\ \bibinfo {author} {\bibfnamefont {G.~J.}\ \bibnamefont
  {Milburn}},\ }\bibfield  {title} {\bibinfo {title} {Quantum theory of
  field-quadrature measurements},\ }\href
  {https://doi.org/10.1103/PhysRevA.47.642} {\bibfield  {journal} {\bibinfo
  {journal} {Physical Review A}\ }\textbf {\bibinfo {volume} {47}},\ \bibinfo
  {pages} {642} (\bibinfo {year} {1993})}\BibitemShut {NoStop}%
\bibitem [{\citenamefont {Jacobs}(2014)}]{jacobsQuantumMeasurementTheory2014}%
  \BibitemOpen
  \bibfield  {author} {\bibinfo {author} {\bibfnamefont {K.}~\bibnamefont
  {Jacobs}},\ }\href {https://doi.org/10.1017/CBO9781139179027} {\emph
  {\bibinfo {title} {Quantum {{Measurement Theory}} and Its
  {{Applications}}}}}\ (\bibinfo  {publisher} {Cambridge University Press},\
  \bibinfo {address} {Cambridge},\ \bibinfo {year} {2014})\BibitemShut
  {NoStop}%
\bibitem [{\citenamefont {Cox}\ \emph {et~al.}(2016)\citenamefont {Cox},
  \citenamefont {Greve}, \citenamefont {Weiner},\ and\ \citenamefont
  {Thompson}}]{coxDeterministicSqueezedStates2016}%
  \BibitemOpen
  \bibfield  {author} {\bibinfo {author} {\bibfnamefont {K.~C.}\ \bibnamefont
  {Cox}}, \bibinfo {author} {\bibfnamefont {G.~P.}\ \bibnamefont {Greve}},
  \bibinfo {author} {\bibfnamefont {J.~M.}\ \bibnamefont {Weiner}},\ and\
  \bibinfo {author} {\bibfnamefont {J.~K.}\ \bibnamefont {Thompson}},\
  }\bibfield  {title} {\bibinfo {title} {Deterministic {{Squeezed States}} with
  {{Collective Measurements}} and {{Feedback}}},\ }\href
  {https://doi.org/10.1103/PhysRevLett.116.093602} {\bibfield  {journal}
  {\bibinfo  {journal} {Physical Review Letters}\ }\textbf {\bibinfo {volume}
  {116}},\ \bibinfo {pages} {093602} (\bibinfo {year} {2016})}\BibitemShut
  {NoStop}%
\bibitem [{\citenamefont {Thomsen}\ \emph {et~al.}(2002)\citenamefont
  {Thomsen}, \citenamefont {Mancini},\ and\ \citenamefont
  {Wiseman}}]{thomsenSpinSqueezingQuantum2002}%
  \BibitemOpen
  \bibfield  {author} {\bibinfo {author} {\bibfnamefont {L.~K.}\ \bibnamefont
  {Thomsen}}, \bibinfo {author} {\bibfnamefont {S.}~\bibnamefont {Mancini}},\
  and\ \bibinfo {author} {\bibfnamefont {H.~M.}\ \bibnamefont {Wiseman}},\
  }\bibfield  {title} {\bibinfo {title} {Spin squeezing via quantum feedback},\
  }\href {https://doi.org/10.1103/PhysRevA.65.061801} {\bibfield  {journal}
  {\bibinfo  {journal} {Physical Review A}\ }\textbf {\bibinfo {volume} {65}},\
  \bibinfo {pages} {061801} (\bibinfo {year} {2002})}\BibitemShut {NoStop}%
\bibitem [{\citenamefont {Inoue}\ \emph {et~al.}(2013)\citenamefont {Inoue},
  \citenamefont {Tanaka}, \citenamefont {Namiki}, \citenamefont {Sagawa},\ and\
  \citenamefont {Takahashi}}]{inoueUnconditionalQuantumNoiseSuppression2013}%
  \BibitemOpen
  \bibfield  {author} {\bibinfo {author} {\bibfnamefont {R.}~\bibnamefont
  {Inoue}}, \bibinfo {author} {\bibfnamefont {S.-I.-R.}\ \bibnamefont
  {Tanaka}}, \bibinfo {author} {\bibfnamefont {R.}~\bibnamefont {Namiki}},
  \bibinfo {author} {\bibfnamefont {T.}~\bibnamefont {Sagawa}},\ and\ \bibinfo
  {author} {\bibfnamefont {Y.}~\bibnamefont {Takahashi}},\ }\bibfield  {title}
  {\bibinfo {title} {Unconditional {{Quantum-Noise Suppression}} via
  {{Measurement-Based Quantum Feedback}}},\ }\href
  {https://doi.org/10.1103/PhysRevLett.110.163602} {\bibfield  {journal}
  {\bibinfo  {journal} {Physical Review Letters}\ }\textbf {\bibinfo {volume}
  {110}},\ \bibinfo {pages} {163602} (\bibinfo {year} {2013})}\BibitemShut
  {NoStop}%
\bibitem [{\citenamefont {Hosten}\ \emph {et~al.}(2016)\citenamefont {Hosten},
  \citenamefont {Engelsen}, \citenamefont {Krishnakumar},\ and\ \citenamefont
  {Kasevich}}]{hostenMeasurementNoise1002016}%
  \BibitemOpen
  \bibfield  {author} {\bibinfo {author} {\bibfnamefont {O.}~\bibnamefont
  {Hosten}}, \bibinfo {author} {\bibfnamefont {N.~J.}\ \bibnamefont
  {Engelsen}}, \bibinfo {author} {\bibfnamefont {R.}~\bibnamefont
  {Krishnakumar}},\ and\ \bibinfo {author} {\bibfnamefont {M.~A.}\ \bibnamefont
  {Kasevich}},\ }\bibfield  {title} {\bibinfo {title} {Measurement noise 100
  times lower than the quantum-projection limit using entangled atoms},\ }\href
  {https://doi.org/10.1038/nature16176} {\bibfield  {journal} {\bibinfo
  {journal} {Nature}\ }\textbf {\bibinfo {volume} {529}},\ \bibinfo {pages}
  {505} (\bibinfo {year} {2016})}\BibitemShut {NoStop}%
\bibitem [{\citenamefont {Guo}\ \emph {et~al.}(2019)\citenamefont {Guo},
  \citenamefont {Norte},\ and\ \citenamefont
  {Gr{\"o}blacher}}]{guoFeedbackCoolingRoom2019}%
  \BibitemOpen
  \bibfield  {author} {\bibinfo {author} {\bibfnamefont {J.}~\bibnamefont
  {Guo}}, \bibinfo {author} {\bibfnamefont {R.}~\bibnamefont {Norte}},\ and\
  \bibinfo {author} {\bibfnamefont {S.}~\bibnamefont {Gr{\"o}blacher}},\
  }\bibfield  {title} {\bibinfo {title} {Feedback {{Cooling}} of a {{Room
  Temperature Mechanical Oscillator}} close to its {{Motional Ground State}}},\
  }\href {https://doi.org/10.1103/PhysRevLett.123.223602} {\bibfield  {journal}
  {\bibinfo  {journal} {Physical Review Letters}\ }\textbf {\bibinfo {volume}
  {123}},\ \bibinfo {pages} {223602} (\bibinfo {year} {2019})}\BibitemShut
  {NoStop}%
\bibitem [{\citenamefont {Sch{\"a}fermeier}\ \emph {et~al.}(2016)\citenamefont
  {Sch{\"a}fermeier}, \citenamefont {Kerdoncuff}, \citenamefont {Hoff},
  \citenamefont {Fu}, \citenamefont {Huck}, \citenamefont {Bilek},
  \citenamefont {Harris}, \citenamefont {Bowen}, \citenamefont {Gehring},\ and\
  \citenamefont {Andersen}}]{schafermeierQuantumEnhancedFeedback2016}%
  \BibitemOpen
  \bibfield  {author} {\bibinfo {author} {\bibfnamefont {C.}~\bibnamefont
  {Sch{\"a}fermeier}}, \bibinfo {author} {\bibfnamefont {H.}~\bibnamefont
  {Kerdoncuff}}, \bibinfo {author} {\bibfnamefont {U.~B.}\ \bibnamefont
  {Hoff}}, \bibinfo {author} {\bibfnamefont {H.}~\bibnamefont {Fu}}, \bibinfo
  {author} {\bibfnamefont {A.}~\bibnamefont {Huck}}, \bibinfo {author}
  {\bibfnamefont {J.}~\bibnamefont {Bilek}}, \bibinfo {author} {\bibfnamefont
  {G.~I.}\ \bibnamefont {Harris}}, \bibinfo {author} {\bibfnamefont {W.~P.}\
  \bibnamefont {Bowen}}, \bibinfo {author} {\bibfnamefont {T.}~\bibnamefont
  {Gehring}},\ and\ \bibinfo {author} {\bibfnamefont {U.~L.}\ \bibnamefont
  {Andersen}},\ }\bibfield  {title} {\bibinfo {title} {Quantum enhanced
  feedback cooling of a mechanical oscillator using nonclassical light},\
  }\href {https://doi.org/10.1038/ncomms13628} {\bibfield  {journal} {\bibinfo
  {journal} {Nature Communications}\ }\textbf {\bibinfo {volume} {7}},\
  \bibinfo {pages} {13628} (\bibinfo {year} {2016})}\BibitemShut {NoStop}%
\bibitem [{\citenamefont {Gajdacz}\ \emph {et~al.}(2016)\citenamefont
  {Gajdacz}, \citenamefont {Hilliard}, \citenamefont {Kristensen},
  \citenamefont {Pedersen}, \citenamefont {Klempt}, \citenamefont {Arlt},\ and\
  \citenamefont {Sherson}}]{gajdaczPreparationUltracoldAtom2016}%
  \BibitemOpen
  \bibfield  {author} {\bibinfo {author} {\bibfnamefont {M.}~\bibnamefont
  {Gajdacz}}, \bibinfo {author} {\bibfnamefont {A.~J.}\ \bibnamefont
  {Hilliard}}, \bibinfo {author} {\bibfnamefont {M.~A.}\ \bibnamefont
  {Kristensen}}, \bibinfo {author} {\bibfnamefont {P.~L.}\ \bibnamefont
  {Pedersen}}, \bibinfo {author} {\bibfnamefont {C.}~\bibnamefont {Klempt}},
  \bibinfo {author} {\bibfnamefont {J.~J.}\ \bibnamefont {Arlt}},\ and\
  \bibinfo {author} {\bibfnamefont {J.~F.}\ \bibnamefont {Sherson}},\
  }\bibfield  {title} {\bibinfo {title} {Preparation of {{Ultracold Atom
  Clouds}} at the {{Shot Noise Level}}},\ }\href
  {https://doi.org/10.1103/PhysRevLett.117.073604} {\bibfield  {journal}
  {\bibinfo  {journal} {Physical Review Letters}\ }\textbf {\bibinfo {volume}
  {117}},\ \bibinfo {pages} {073604} (\bibinfo {year} {2016})}\BibitemShut
  {NoStop}%
\bibitem [{\citenamefont {Haine}\ \emph {et~al.}(2004)\citenamefont {Haine},
  \citenamefont {Ferris}, \citenamefont {Close},\ and\ \citenamefont
  {Hope}}]{hainefeedback}%
  \BibitemOpen
  \bibfield  {author} {\bibinfo {author} {\bibfnamefont {S.~A.}\ \bibnamefont
  {Haine}}, \bibinfo {author} {\bibfnamefont {A.~J.}\ \bibnamefont {Ferris}},
  \bibinfo {author} {\bibfnamefont {J.~D.}\ \bibnamefont {Close}},\ and\
  \bibinfo {author} {\bibfnamefont {J.~J.}\ \bibnamefont {Hope}},\ }\bibfield
  {title} {\bibinfo {title} {Control of an atom laser using feedback},\ }\href
  {https://doi.org/10.1103/PhysRevA.69.013605} {\bibfield  {journal} {\bibinfo
  {journal} {Physical Review A: Atomic, Molecular, and Optical Physics}\
  }\textbf {\bibinfo {volume} {69}},\ \bibinfo {pages} {013605} (\bibinfo
  {year} {2004})}\BibitemShut {NoStop}%
\bibitem [{\citenamefont {Szigeti}\ \emph {et~al.}(2009)\citenamefont
  {Szigeti}, \citenamefont {Hush}, \citenamefont {Carvalho},\ and\
  \citenamefont {Hope}}]{szigetiphasecontrast}%
  \BibitemOpen
  \bibfield  {author} {\bibinfo {author} {\bibfnamefont {S.~S.}\ \bibnamefont
  {Szigeti}}, \bibinfo {author} {\bibfnamefont {M.~R.}\ \bibnamefont {Hush}},
  \bibinfo {author} {\bibfnamefont {A.~R.~R.}\ \bibnamefont {Carvalho}},\ and\
  \bibinfo {author} {\bibfnamefont {J.~J.}\ \bibnamefont {Hope}},\ }\bibfield
  {title} {\bibinfo {title} {Continuous measurement feedback control of a
  {{Bose-Einstein}} condensate using phase-contrast imaging},\ }\href
  {https://doi.org/10.1103/PhysRevA.80.013614} {\bibfield  {journal} {\bibinfo
  {journal} {Physical Review A: Atomic, Molecular, and Optical Physics}\
  }\textbf {\bibinfo {volume} {80}},\ \bibinfo {pages} {013614} (\bibinfo
  {year} {2009})}\BibitemShut {NoStop}%
\bibitem [{\citenamefont {Hush}\ \emph {et~al.}(2013)\citenamefont {Hush},
  \citenamefont {Szigeti}, \citenamefont {Carvalho},\ and\ \citenamefont
  {Hope}}]{hushControllingSpontaneousemissionNoise2013}%
  \BibitemOpen
  \bibfield  {author} {\bibinfo {author} {\bibfnamefont {M.~R.}\ \bibnamefont
  {Hush}}, \bibinfo {author} {\bibfnamefont {S.~S.}\ \bibnamefont {Szigeti}},
  \bibinfo {author} {\bibfnamefont {A.~R.~R.}\ \bibnamefont {Carvalho}},\ and\
  \bibinfo {author} {\bibfnamefont {J.~J.}\ \bibnamefont {Hope}},\ }\bibfield
  {title} {\bibinfo {title} {Controlling spontaneous-emission noise in
  measurement-based feedback cooling of a {{Bose}}--{{Einstein}} condensate},\
  }\href {https://doi.org/10.1088/1367-2630/15/11/113060} {\bibfield  {journal}
  {\bibinfo  {journal} {New Journal of Physics}\ }\textbf {\bibinfo {volume}
  {15}},\ \bibinfo {pages} {113060} (\bibinfo {year} {2013})}\BibitemShut
  {NoStop}%
\bibitem [{\citenamefont {Goh}\ \emph {et~al.}(2022)\citenamefont {Goh},
  \citenamefont {Mehdi}, \citenamefont {Taylor}, \citenamefont {Thomas},
  \citenamefont {Bradley}, \citenamefont {Hush}, \citenamefont {Hope},\ and\
  \citenamefont {Szigeti}}]{gohFeedbackCoolingBose2022}%
  \BibitemOpen
  \bibfield  {author} {\bibinfo {author} {\bibfnamefont {M.~L.}\ \bibnamefont
  {Goh}}, \bibinfo {author} {\bibfnamefont {Z.}~\bibnamefont {Mehdi}}, \bibinfo
  {author} {\bibfnamefont {R.~L.}\ \bibnamefont {Taylor}}, \bibinfo {author}
  {\bibfnamefont {R.~J.}\ \bibnamefont {Thomas}}, \bibinfo {author}
  {\bibfnamefont {A.~S.}\ \bibnamefont {Bradley}}, \bibinfo {author}
  {\bibfnamefont {M.~R.}\ \bibnamefont {Hush}}, \bibinfo {author}
  {\bibfnamefont {J.~J.}\ \bibnamefont {Hope}},\ and\ \bibinfo {author}
  {\bibfnamefont {S.~S.}\ \bibnamefont {Szigeti}},\ }\href
  {https://doi.org/10.48550/arXiv.2206.05069} {\bibinfo {title} {Feedback
  cooling {{Bose}} gases to quantum degeneracy}} (\bibinfo {year} {2022}),\
  \Eprint {https://arxiv.org/abs/2206.05069} {arXiv:2206.05069 [cond-mat,
  physics:quant-ph]} \BibitemShut {NoStop}%
\bibitem [{\citenamefont {Hurst}\ \emph {et~al.}(2020)\citenamefont {Hurst},
  \citenamefont {Guo},\ and\ \citenamefont
  {Spielman}}]{hurstFeedbackInducedMagnetic2020}%
  \BibitemOpen
  \bibfield  {author} {\bibinfo {author} {\bibfnamefont {H.~M.}\ \bibnamefont
  {Hurst}}, \bibinfo {author} {\bibfnamefont {S.}~\bibnamefont {Guo}},\ and\
  \bibinfo {author} {\bibfnamefont {I.~B.}\ \bibnamefont {Spielman}},\
  }\bibfield  {title} {\bibinfo {title} {Feedback induced magnetic phases in
  binary {{Bose-Einstein}} condensates},\ }\href
  {https://doi.org/10.1103/PhysRevResearch.2.043325} {\bibfield  {journal}
  {\bibinfo  {journal} {Physical Review Research}\ }\textbf {\bibinfo {volume}
  {2}},\ \bibinfo {pages} {043325} (\bibinfo {year} {2020})}\BibitemShut
  {NoStop}%
\bibitem [{\citenamefont {Yamaguchi}\ \emph {et~al.}(2023)\citenamefont
  {Yamaguchi}, \citenamefont {Hurst},\ and\ \citenamefont
  {Spielman}}]{yamaguchiFeedbackcooledBoseEinsteinCondensation2023}%
  \BibitemOpen
  \bibfield  {author} {\bibinfo {author} {\bibfnamefont {E.~P.}\ \bibnamefont
  {Yamaguchi}}, \bibinfo {author} {\bibfnamefont {H.~M.}\ \bibnamefont
  {Hurst}},\ and\ \bibinfo {author} {\bibfnamefont {I.~B.}\ \bibnamefont
  {Spielman}},\ }\bibfield  {title} {\bibinfo {title} {Feedback-cooled
  {{Bose-Einstein}} condensation: {{Near}} and far from equilibrium},\ }\href
  {https://doi.org/10.1103/PhysRevA.107.063306} {\bibfield  {journal} {\bibinfo
   {journal} {Physical Review A}\ }\textbf {\bibinfo {volume} {107}},\ \bibinfo
  {pages} {063306} (\bibinfo {year} {2023})}\BibitemShut {NoStop}%
\bibitem [{\citenamefont {Young}\ \emph {et~al.}(2021)\citenamefont {Young},
  \citenamefont {Gorshkov},\ and\ \citenamefont
  {Spielman}}]{youngFeedbackstabilizedDynamicalSteady2021}%
  \BibitemOpen
  \bibfield  {author} {\bibinfo {author} {\bibfnamefont {J.~T.}\ \bibnamefont
  {Young}}, \bibinfo {author} {\bibfnamefont {A.~V.}\ \bibnamefont
  {Gorshkov}},\ and\ \bibinfo {author} {\bibfnamefont {I.~B.}\ \bibnamefont
  {Spielman}},\ }\bibfield  {title} {\bibinfo {title} {Feedback-stabilized
  dynamical steady states in the {{Bose-Hubbard}} model},\ }\href
  {https://doi.org/10.1103/PhysRevResearch.3.043075} {\bibfield  {journal}
  {\bibinfo  {journal} {Physical Review Research}\ }\textbf {\bibinfo {volume}
  {3}},\ \bibinfo {pages} {043075} (\bibinfo {year} {2021})}\BibitemShut
  {NoStop}%
\bibitem [{\citenamefont {{Mu{\~n}oz-Arias}}\ \emph {et~al.}(2020)\citenamefont
  {{Mu{\~n}oz-Arias}}, \citenamefont {Poggi}, \citenamefont {Jessen},\ and\
  \citenamefont {Deutsch}}]{munozariasSimulatingNonlinearDynamics2020a}%
  \BibitemOpen
  \bibfield  {author} {\bibinfo {author} {\bibfnamefont {M.~H.}\ \bibnamefont
  {{Mu{\~n}oz-Arias}}}, \bibinfo {author} {\bibfnamefont {P.~M.}\ \bibnamefont
  {Poggi}}, \bibinfo {author} {\bibfnamefont {P.~S.}\ \bibnamefont {Jessen}},\
  and\ \bibinfo {author} {\bibfnamefont {I.~H.}\ \bibnamefont {Deutsch}},\
  }\bibfield  {title} {\bibinfo {title} {Simulating {{Nonlinear Dynamics}} of
  {{Collective Spins}} via {{Quantum Measurement}} and {{Feedback}}},\ }\href
  {https://doi.org/10.1103/PhysRevLett.124.110503} {\bibfield  {journal}
  {\bibinfo  {journal} {Physical Review Letters}\ }\textbf {\bibinfo {volume}
  {124}},\ \bibinfo {pages} {110503} (\bibinfo {year} {2020})}\BibitemShut
  {NoStop}%
\bibitem [{\citenamefont {Mehdi}\ \emph {et~al.}(2024)\citenamefont {Mehdi},
  \citenamefont {Haine}, \citenamefont {Hope},\ and\ \citenamefont
  {Szigeti}}]{mehdiFundamentalLimitsFeedback2024}%
  \BibitemOpen
  \bibfield  {author} {\bibinfo {author} {\bibfnamefont {Z.}~\bibnamefont
  {Mehdi}}, \bibinfo {author} {\bibfnamefont {S.~A.}\ \bibnamefont {Haine}},
  \bibinfo {author} {\bibfnamefont {J.~J.}\ \bibnamefont {Hope}},\ and\
  \bibinfo {author} {\bibfnamefont {S.~S.}\ \bibnamefont {Szigeti}},\
  }\bibfield  {title} {\bibinfo {title} {Fundamental {{Limits}} of {{Feedback
  Cooling Ultracold Atomic Gases}}},\ }\href
  {https://doi.org/10.1103/PhysRevLett.133.073401} {\bibfield  {journal}
  {\bibinfo  {journal} {Physical Review Letters}\ }\textbf {\bibinfo {volume}
  {133}},\ \bibinfo {pages} {073401} (\bibinfo {year} {2024})}\BibitemShut
  {NoStop}%
\bibitem [{\citenamefont {Blakie{\dag}}\ \emph {et~al.}(2008)\citenamefont
  {Blakie{\dag}}, \citenamefont {Bradley{\dag}}, \citenamefont {Davis},
  \citenamefont {Ballagh},\ and\ \citenamefont {{C.W. Gardiner}}}]{psmethods}%
  \BibitemOpen
  \bibfield  {author} {\bibinfo {author} {\bibfnamefont {P.}~\bibnamefont
  {Blakie{\dag}}}, \bibinfo {author} {\bibfnamefont {A.}~\bibnamefont
  {Bradley{\dag}}}, \bibinfo {author} {\bibfnamefont {M.}~\bibnamefont
  {Davis}}, \bibinfo {author} {\bibfnamefont {R.}~\bibnamefont {Ballagh}},\
  and\ \bibinfo {author} {\bibnamefont {{C.W. Gardiner}}},\ }\bibfield  {title}
  {\bibinfo {title} {Dynamics and statistical mechanics of ultra-cold {{Bose}}
  gases using c-field techniques},\ }\href
  {https://doi.org/10.1080/00018730802564254} {\bibfield  {journal} {\bibinfo
  {journal} {Advances in Physics}\ }\textbf {\bibinfo {volume} {57}},\ \bibinfo
  {pages} {363} (\bibinfo {year} {2008})},\ \Eprint
  {https://arxiv.org/abs/https://doi.org/10.1080/00018730802564254}
  {https://doi.org/10.1080/00018730802564254} \BibitemShut {NoStop}%
\bibitem [{\citenamefont {Gardiner}\ and\ \citenamefont
  {Zoller}(2010)}]{gardiner}%
  \BibitemOpen
  \bibfield  {author} {\bibinfo {author} {\bibfnamefont {C.~W.}\ \bibnamefont
  {Gardiner}}\ and\ \bibinfo {author} {\bibfnamefont {P.}~\bibnamefont
  {Zoller}},\ }\href@noop {} {\emph {\bibinfo {title} {Quantum Noise}}},\
  Springer Series in Synergetics\ (\bibinfo  {publisher} {Springer},\ \bibinfo
  {address} {Berlin, Germany},\ \bibinfo {year} {2010})\BibitemShut {NoStop}%
\bibitem [{\citenamefont {Behbood}\ \emph {et~al.}(2013)\citenamefont
  {Behbood}, \citenamefont {Colangelo}, \citenamefont {Martin~Ciurana},
  \citenamefont {Napolitano}, \citenamefont {Sewell},\ and\ \citenamefont
  {Mitchell}}]{behboodFeedbackCoolingAtomic2013}%
  \BibitemOpen
  \bibfield  {author} {\bibinfo {author} {\bibfnamefont {N.}~\bibnamefont
  {Behbood}}, \bibinfo {author} {\bibfnamefont {G.}~\bibnamefont {Colangelo}},
  \bibinfo {author} {\bibfnamefont {F.}~\bibnamefont {Martin~Ciurana}},
  \bibinfo {author} {\bibfnamefont {M.}~\bibnamefont {Napolitano}}, \bibinfo
  {author} {\bibfnamefont {R.~J.}\ \bibnamefont {Sewell}},\ and\ \bibinfo
  {author} {\bibfnamefont {M.~W.}\ \bibnamefont {Mitchell}},\ }\bibfield
  {title} {\bibinfo {title} {Feedback {{Cooling}} of an {{Atomic Spin
  Ensemble}}},\ }\href {https://doi.org/10.1103/PhysRevLett.111.103601}
  {\bibfield  {journal} {\bibinfo  {journal} {Physical Review Letters}\
  }\textbf {\bibinfo {volume} {111}},\ \bibinfo {pages} {103601} (\bibinfo
  {year} {2013})}\BibitemShut {NoStop}%
\bibitem [{\citenamefont {Odelli}\ \emph {et~al.}(2024)\citenamefont {Odelli},
  \citenamefont {Ruschhaupt},\ and\ \citenamefont
  {Stojanovi{\'c}}}]{odelliTwistandturnDynamicsSpin2024}%
  \BibitemOpen
  \bibfield  {author} {\bibinfo {author} {\bibfnamefont {M.}~\bibnamefont
  {Odelli}}, \bibinfo {author} {\bibfnamefont {A.}~\bibnamefont {Ruschhaupt}},\
  and\ \bibinfo {author} {\bibfnamefont {V.~M.}\ \bibnamefont
  {Stojanovi{\'c}}},\ }\bibfield  {title} {\bibinfo {title} {Twist-and-turn
  dynamics of spin squeezing in bosonic {{Josephson}} junctions: {{Enhanced}}
  shortcuts-to-adiabaticity approach},\ }\href
  {https://doi.org/10.1103/PhysRevA.110.022610} {\bibfield  {journal} {\bibinfo
   {journal} {Physical Review A}\ }\textbf {\bibinfo {volume} {110}},\ \bibinfo
  {pages} {022610} (\bibinfo {year} {2024})}\BibitemShut {NoStop}%
\bibitem [{\citenamefont {Wiseman}\ and\ \citenamefont
  {Milburn}(2009)}]{wisemanQuantumMeasurementControl2009}%
  \BibitemOpen
  \bibfield  {author} {\bibinfo {author} {\bibfnamefont {H.~M.}\ \bibnamefont
  {Wiseman}}\ and\ \bibinfo {author} {\bibfnamefont {G.~J.}\ \bibnamefont
  {Milburn}},\ }\href {https://doi.org/10.1017/CBO9780511813948} {\emph
  {\bibinfo {title} {Quantum {{Measurement}} and {{Control}}}}}\ (\bibinfo
  {publisher} {Cambridge University Press},\ \bibinfo {address} {Cambridge},\
  \bibinfo {year} {2009})\BibitemShut {NoStop}%
\bibitem [{\citenamefont {Rundle}\ and\ \citenamefont
  {Everitt}(2021)}]{rundleOverviewPhaseSpace2021}%
  \BibitemOpen
  \bibfield  {author} {\bibinfo {author} {\bibfnamefont {R.~P.}\ \bibnamefont
  {Rundle}}\ and\ \bibinfo {author} {\bibfnamefont {M.~J.}\ \bibnamefont
  {Everitt}},\ }\bibfield  {title} {\bibinfo {title} {Overview of the {{Phase
  Space Formulation}} of {{Quantum Mechanics}} with {{Application}} to
  {{Quantum Technologies}}},\ }\href {https://doi.org/10.1002/qute.202100016}
  {\bibfield  {journal} {\bibinfo  {journal} {Advanced Quantum Technologies}\
  }\textbf {\bibinfo {volume} {4}},\ \bibinfo {pages} {2100016} (\bibinfo
  {year} {2021})}\BibitemShut {NoStop}%
\bibitem [{\citenamefont {Lee}(1995)}]{leeTheoryApplicationQuantum1995}%
  \BibitemOpen
  \bibfield  {author} {\bibinfo {author} {\bibfnamefont {H.-W.}\ \bibnamefont
  {Lee}},\ }\bibfield  {title} {\bibinfo {title} {Theory and application of the
  quantum phase-space distribution functions},\ }\href
  {https://doi.org/10.1016/0370-1573(95)00007-4} {\bibfield  {journal}
  {\bibinfo  {journal} {Physics Reports}\ }\textbf {\bibinfo {volume} {259}},\
  \bibinfo {pages} {147} (\bibinfo {year} {1995})}\BibitemShut {NoStop}%
\bibitem [{\citenamefont
  {Gardiner}(1985)}]{gardinerHandbookStochasticMethods1985}%
  \BibitemOpen
  \bibfield  {author} {\bibinfo {author} {\bibfnamefont {C.~W.}\ \bibnamefont
  {Gardiner}},\ }\href@noop {} {\emph {\bibinfo {title} {Handbook of
  {{Stochastic Methods}} for {{Physics}}, {{Chemistry}}, and the {{Natural
  Sciences}}}}}\ (\bibinfo  {publisher} {Springer-Verlag},\ \bibinfo {year}
  {1985})\BibitemShut {NoStop}%
\bibitem [{\citenamefont
  {Wigner}(1932)}]{wignerQuantumCorrectionThermodynamic1932}%
  \BibitemOpen
  \bibfield  {author} {\bibinfo {author} {\bibfnamefont {E.}~\bibnamefont
  {Wigner}},\ }\bibfield  {title} {\bibinfo {title} {On the {{Quantum
  Correction For Thermodynamic Equilibrium}}},\ }\href
  {https://doi.org/10.1103/PhysRev.40.749} {\bibfield  {journal} {\bibinfo
  {journal} {Physical Review}\ }\textbf {\bibinfo {volume} {40}},\ \bibinfo
  {pages} {749} (\bibinfo {year} {1932})}\BibitemShut {NoStop}%
\bibitem [{\citenamefont {Drummond}\ and\ \citenamefont
  {Opanchuk}(2017{\natexlab{a}})}]{drummondTruncatedWignerDynamics2017a}%
  \BibitemOpen
  \bibfield  {author} {\bibinfo {author} {\bibfnamefont {P.~D.}\ \bibnamefont
  {Drummond}}\ and\ \bibinfo {author} {\bibfnamefont {B.}~\bibnamefont
  {Opanchuk}},\ }\bibfield  {title} {\bibinfo {title} {Truncated {{Wigner}}
  dynamics and conservation laws},\ }\href
  {https://doi.org/10.1103/PhysRevA.96.043616} {\bibfield  {journal} {\bibinfo
  {journal} {Physical Review A}\ }\textbf {\bibinfo {volume} {96}},\ \bibinfo
  {pages} {043616} (\bibinfo {year} {2017}{\natexlab{a}})}\BibitemShut
  {NoStop}%
\bibitem [{\citenamefont
  {Polkovnikov}(2010)}]{polkovnikovPhaseSpaceRepresentation2010}%
  \BibitemOpen
  \bibfield  {author} {\bibinfo {author} {\bibfnamefont {A.}~\bibnamefont
  {Polkovnikov}},\ }\bibfield  {title} {\bibinfo {title} {Phase space
  representation of quantum dynamics},\ }\href
  {https://doi.org/10.1016/j.aop.2010.02.006} {\bibfield  {journal} {\bibinfo
  {journal} {Annals of Physics}\ }\textbf {\bibinfo {volume} {325}},\ \bibinfo
  {pages} {1790} (\bibinfo {year} {2010})},\ \Eprint
  {https://arxiv.org/abs/0905.3384} {arXiv:0905.3384 [cond-mat, physics:hep-th,
  physics:quant-ph]} \BibitemShut {NoStop}%
\bibitem [{\citenamefont {Steel}\ \emph {et~al.}(1998)\citenamefont {Steel},
  \citenamefont {Olsen}, \citenamefont {Plimak}, \citenamefont {Drummond},
  \citenamefont {Tan}, \citenamefont {Collett}, \citenamefont {Walls},\ and\
  \citenamefont {Graham}}]{steelDynamicalQuantumNoise1998}%
  \BibitemOpen
  \bibfield  {author} {\bibinfo {author} {\bibfnamefont {M.~J.}\ \bibnamefont
  {Steel}}, \bibinfo {author} {\bibfnamefont {M.~K.}\ \bibnamefont {Olsen}},
  \bibinfo {author} {\bibfnamefont {L.~I.}\ \bibnamefont {Plimak}}, \bibinfo
  {author} {\bibfnamefont {P.~D.}\ \bibnamefont {Drummond}}, \bibinfo {author}
  {\bibfnamefont {S.~M.}\ \bibnamefont {Tan}}, \bibinfo {author} {\bibfnamefont
  {M.~J.}\ \bibnamefont {Collett}}, \bibinfo {author} {\bibfnamefont {D.~F.}\
  \bibnamefont {Walls}},\ and\ \bibinfo {author} {\bibfnamefont
  {R.}~\bibnamefont {Graham}},\ }\bibfield  {title} {\bibinfo {title}
  {Dynamical quantum noise in trapped {{Bose-Einstein}} condensates},\ }\href
  {https://doi.org/10.1103/PhysRevA.58.4824} {\bibfield  {journal} {\bibinfo
  {journal} {Physical Review A}\ }\textbf {\bibinfo {volume} {58}},\ \bibinfo
  {pages} {4824} (\bibinfo {year} {1998})}\BibitemShut {NoStop}%
\bibitem [{\citenamefont {Sinatra}\ \emph {et~al.}(2002)\citenamefont
  {Sinatra}, \citenamefont {Lobo},\ and\ \citenamefont
  {Castin}}]{sinatraTruncatedWignerMethod2002}%
  \BibitemOpen
  \bibfield  {author} {\bibinfo {author} {\bibfnamefont {A.}~\bibnamefont
  {Sinatra}}, \bibinfo {author} {\bibfnamefont {C.}~\bibnamefont {Lobo}},\ and\
  \bibinfo {author} {\bibfnamefont {Y.}~\bibnamefont {Castin}},\ }\bibfield
  {title} {\bibinfo {title} {The truncated {{Wigner}} method for
  {{Bose-condensed}} gases: Limits of validity and applications1},\ }\href
  {https://doi.org/10.1088/0953-4075/35/17/301} {\bibfield  {journal} {\bibinfo
   {journal} {Journal of Physics B: Atomic, Molecular and Optical Physics}\
  }\textbf {\bibinfo {volume} {35}},\ \bibinfo {pages} {3599} (\bibinfo {year}
  {2002})}\BibitemShut {NoStop}%
\bibitem [{\citenamefont {Drummond}\ and\ \citenamefont
  {Opanchuk}(2017{\natexlab{b}})}]{drummondTruncatedWignerDynamics2017}%
  \BibitemOpen
  \bibfield  {author} {\bibinfo {author} {\bibfnamefont {P.~D.}\ \bibnamefont
  {Drummond}}\ and\ \bibinfo {author} {\bibfnamefont {B.}~\bibnamefont
  {Opanchuk}},\ }\bibfield  {title} {\bibinfo {title} {Truncated {{Wigner}}
  dynamics and conservation laws},\ }\href
  {https://doi.org/10.1103/PhysRevA.96.043616} {\bibfield  {journal} {\bibinfo
  {journal} {Physical Review A}\ }\textbf {\bibinfo {volume} {96}},\ \bibinfo
  {pages} {043616} (\bibinfo {year} {2017}{\natexlab{b}})}\BibitemShut
  {NoStop}%
\bibitem [{\citenamefont {Haine}(2018)}]{haineQuantumNoiseBright2018}%
  \BibitemOpen
  \bibfield  {author} {\bibinfo {author} {\bibfnamefont {S.~A.}\ \bibnamefont
  {Haine}},\ }\bibfield  {title} {\bibinfo {title} {Quantum noise in bright
  soliton matterwave interferometry},\ }\href
  {https://doi.org/10.1088/1367-2630/aab47f} {\bibfield  {journal} {\bibinfo
  {journal} {New Journal of Physics}\ }\textbf {\bibinfo {volume} {20}},\
  \bibinfo {pages} {033009} (\bibinfo {year} {2018})}\BibitemShut {NoStop}%
\bibitem [{\citenamefont {Norrie}\ \emph {et~al.}(2006)\citenamefont {Norrie},
  \citenamefont {Ballagh},\ and\ \citenamefont
  {Gardiner}}]{norrieQuantumTurbulenceCorrelations2006}%
  \BibitemOpen
  \bibfield  {author} {\bibinfo {author} {\bibfnamefont {A.~A.}\ \bibnamefont
  {Norrie}}, \bibinfo {author} {\bibfnamefont {R.~J.}\ \bibnamefont
  {Ballagh}},\ and\ \bibinfo {author} {\bibfnamefont {C.~W.}\ \bibnamefont
  {Gardiner}},\ }\bibfield  {title} {\bibinfo {title} {Quantum turbulence and
  correlations in {{Bose-Einstein}} condensate collisions},\ }\href
  {https://doi.org/10.1103/PhysRevA.73.043617} {\bibfield  {journal} {\bibinfo
  {journal} {Physical Review A}\ }\textbf {\bibinfo {volume} {73}},\ \bibinfo
  {pages} {043617} (\bibinfo {year} {2006})}\BibitemShut {NoStop}%
\bibitem [{\citenamefont {Haine}\ and\ \citenamefont
  {Ferris}(2011)}]{haineSurpassingStandardQuantum2011}%
  \BibitemOpen
  \bibfield  {author} {\bibinfo {author} {\bibfnamefont {S.~A.}\ \bibnamefont
  {Haine}}\ and\ \bibinfo {author} {\bibfnamefont {A.~J.}\ \bibnamefont
  {Ferris}},\ }\bibfield  {title} {\bibinfo {title} {Surpassing the standard
  quantum limit in an atom interferometer with four-mode entanglement produced
  from four-wave mixing},\ }\href {https://doi.org/10.1103/PhysRevA.84.043624}
  {\bibfield  {journal} {\bibinfo  {journal} {Physical Review A}\ }\textbf
  {\bibinfo {volume} {84}},\ \bibinfo {pages} {043624} (\bibinfo {year}
  {2011})}\BibitemShut {NoStop}%
\bibitem [{\citenamefont {Drummond}\ and\ \citenamefont
  {Hardman}(1993)}]{drummondSimulationQuantumEffects1993}%
  \BibitemOpen
  \bibfield  {author} {\bibinfo {author} {\bibfnamefont {P.~D.}\ \bibnamefont
  {Drummond}}\ and\ \bibinfo {author} {\bibfnamefont {A.~D.}\ \bibnamefont
  {Hardman}},\ }\bibfield  {title} {\bibinfo {title} {Simulation of {{Quantum
  Effects}} in {{Raman-Active Waveguides}}},\ }\href
  {https://doi.org/10.1209/0295-5075/21/3/005} {\bibfield  {journal} {\bibinfo
  {journal} {Europhysics Letters}\ }\textbf {\bibinfo {volume} {21}},\ \bibinfo
  {pages} {279} (\bibinfo {year} {1993})}\BibitemShut {NoStop}%
\bibitem [{\citenamefont {Haine}\ and\ \citenamefont
  {Johnsson}(2009)}]{haineDynamicSchemeGenerating2009}%
  \BibitemOpen
  \bibfield  {author} {\bibinfo {author} {\bibfnamefont {S.~A.}\ \bibnamefont
  {Haine}}\ and\ \bibinfo {author} {\bibfnamefont {M.~T.}\ \bibnamefont
  {Johnsson}},\ }\bibfield  {title} {\bibinfo {title} {Dynamic scheme for
  generating number squeezing in {{Bose-Einstein}} condensates through
  nonlinear interactions},\ }\href {https://doi.org/10.1103/PhysRevA.80.023611}
  {\bibfield  {journal} {\bibinfo  {journal} {Physical Review A}\ }\textbf
  {\bibinfo {volume} {80}},\ \bibinfo {pages} {023611} (\bibinfo {year}
  {2009})}\BibitemShut {NoStop}%
\bibitem [{\citenamefont {Opanchuk}\ \emph {et~al.}(2012)\citenamefont
  {Opanchuk}, \citenamefont {Egorov}, \citenamefont {Hoffmann}, \citenamefont
  {Sidorov},\ and\ \citenamefont
  {Drummond}}]{opanchukQuantumNoiseThreedimensional2012}%
  \BibitemOpen
  \bibfield  {author} {\bibinfo {author} {\bibfnamefont {B.}~\bibnamefont
  {Opanchuk}}, \bibinfo {author} {\bibfnamefont {M.}~\bibnamefont {Egorov}},
  \bibinfo {author} {\bibfnamefont {S.}~\bibnamefont {Hoffmann}}, \bibinfo
  {author} {\bibfnamefont {A.~I.}\ \bibnamefont {Sidorov}},\ and\ \bibinfo
  {author} {\bibfnamefont {P.~D.}\ \bibnamefont {Drummond}},\ }\bibfield
  {title} {\bibinfo {title} {Quantum noise in three-dimensional {{BEC}}
  interferometry},\ }\href {https://doi.org/10.1209/0295-5075/97/50003}
  {\bibfield  {journal} {\bibinfo  {journal} {Europhysics Letters}\ }\textbf
  {\bibinfo {volume} {97}},\ \bibinfo {pages} {50003} (\bibinfo {year}
  {2012})}\BibitemShut {NoStop}%
\bibitem [{\citenamefont {Slodi{\v c}ka}\ \emph {et~al.}(2013)\citenamefont
  {Slodi{\v c}ka}, \citenamefont {H{\'e}tet}, \citenamefont {R{\"o}ck},
  \citenamefont {Schindler}, \citenamefont {Hennrich},\ and\ \citenamefont
  {Blatt}}]{slodickaAtomAtomEntanglementSinglePhoton2013}%
  \BibitemOpen
  \bibfield  {author} {\bibinfo {author} {\bibfnamefont {L.}~\bibnamefont
  {Slodi{\v c}ka}}, \bibinfo {author} {\bibfnamefont {G.}~\bibnamefont
  {H{\'e}tet}}, \bibinfo {author} {\bibfnamefont {N.}~\bibnamefont {R{\"o}ck}},
  \bibinfo {author} {\bibfnamefont {P.}~\bibnamefont {Schindler}}, \bibinfo
  {author} {\bibfnamefont {M.}~\bibnamefont {Hennrich}},\ and\ \bibinfo
  {author} {\bibfnamefont {R.}~\bibnamefont {Blatt}},\ }\bibfield  {title}
  {\bibinfo {title} {Atom-{{Atom Entanglement}} by {{Single-Photon
  Detection}}},\ }\href {https://doi.org/10.1103/PhysRevLett.110.083603}
  {\bibfield  {journal} {\bibinfo  {journal} {Physical Review Letters}\
  }\textbf {\bibinfo {volume} {110}},\ \bibinfo {pages} {083603} (\bibinfo
  {year} {2013})}\BibitemShut {NoStop}%
\bibitem [{\citenamefont {{Lewis-Swan}}\ and\ \citenamefont
  {Kheruntsyan}(2013)}]{lewis-swanSensitivityThermalNoise2013}%
  \BibitemOpen
  \bibfield  {author} {\bibinfo {author} {\bibfnamefont {R.~J.}\ \bibnamefont
  {{Lewis-Swan}}}\ and\ \bibinfo {author} {\bibfnamefont {K.~V.}\ \bibnamefont
  {Kheruntsyan}},\ }\bibfield  {title} {\bibinfo {title} {Sensitivity to
  thermal noise of atomic {{Einstein-Podolsky-Rosen}} entanglement},\ }\href
  {https://doi.org/10.1103/PhysRevA.87.063635} {\bibfield  {journal} {\bibinfo
  {journal} {Physical Review A}\ }\textbf {\bibinfo {volume} {87}},\ \bibinfo
  {pages} {063635} (\bibinfo {year} {2013})}\BibitemShut {NoStop}%
\bibitem [{\citenamefont {Haine}\ \emph {et~al.}(2014)\citenamefont {Haine},
  \citenamefont {Lau}, \citenamefont {Anderson},\ and\ \citenamefont
  {Johnsson}}]{haineSelfinducedSpatialDynamics2014}%
  \BibitemOpen
  \bibfield  {author} {\bibinfo {author} {\bibfnamefont {S.~A.}\ \bibnamefont
  {Haine}}, \bibinfo {author} {\bibfnamefont {J.}~\bibnamefont {Lau}}, \bibinfo
  {author} {\bibfnamefont {R.~P.}\ \bibnamefont {Anderson}},\ and\ \bibinfo
  {author} {\bibfnamefont {M.~T.}\ \bibnamefont {Johnsson}},\ }\bibfield
  {title} {\bibinfo {title} {Self-induced spatial dynamics to enhance spin
  squeezing via one-axis twisting in a two-component {{Bose-Einstein}}
  condensate},\ }\href {https://doi.org/10.1103/PhysRevA.90.023613} {\bibfield
  {journal} {\bibinfo  {journal} {Physical Review A}\ }\textbf {\bibinfo
  {volume} {90}},\ \bibinfo {pages} {023613} (\bibinfo {year}
  {2014})}\BibitemShut {NoStop}%
\bibitem [{\citenamefont {Nolan}\ \emph {et~al.}(2016)\citenamefont {Nolan},
  \citenamefont {Sabbatini}, \citenamefont {Bromley}, \citenamefont {Davis},\
  and\ \citenamefont {Haine}}]{nolanQuantumEnhancedMeasurement2016}%
  \BibitemOpen
  \bibfield  {author} {\bibinfo {author} {\bibfnamefont {S.~P.}\ \bibnamefont
  {Nolan}}, \bibinfo {author} {\bibfnamefont {J.}~\bibnamefont {Sabbatini}},
  \bibinfo {author} {\bibfnamefont {M.~W.~J.}\ \bibnamefont {Bromley}},
  \bibinfo {author} {\bibfnamefont {M.~J.}\ \bibnamefont {Davis}},\ and\
  \bibinfo {author} {\bibfnamefont {S.~A.}\ \bibnamefont {Haine}},\ }\bibfield
  {title} {\bibinfo {title} {Quantum enhanced measurement of rotations with a
  spin-1 {{Bose-Einstein}} condensate in a ring trap},\ }\href
  {https://doi.org/10.1103/PhysRevA.93.023616} {\bibfield  {journal} {\bibinfo
  {journal} {Physical Review A}\ }\textbf {\bibinfo {volume} {93}},\ \bibinfo
  {pages} {023616} (\bibinfo {year} {2016})}\BibitemShut {NoStop}%
\bibitem [{\citenamefont {Szigeti}\ \emph {et~al.}(2020)\citenamefont
  {Szigeti}, \citenamefont {Nolan}, \citenamefont {Close},\ and\ \citenamefont
  {Haine}}]{szigetiHighPrecisionQuantumEnhancedGravimetry2020}%
  \BibitemOpen
  \bibfield  {author} {\bibinfo {author} {\bibfnamefont {S.~S.}\ \bibnamefont
  {Szigeti}}, \bibinfo {author} {\bibfnamefont {S.~P.}\ \bibnamefont {Nolan}},
  \bibinfo {author} {\bibfnamefont {J.~D.}\ \bibnamefont {Close}},\ and\
  \bibinfo {author} {\bibfnamefont {S.~A.}\ \bibnamefont {Haine}},\ }\bibfield
  {title} {\bibinfo {title} {High-{{Precision Quantum-Enhanced Gravimetry}}
  with a {{Bose-Einstein Condensate}}},\ }\href
  {https://doi.org/10.1103/PhysRevLett.125.100402} {\bibfield  {journal}
  {\bibinfo  {journal} {Physical Review Letters}\ }\textbf {\bibinfo {volume}
  {125}},\ \bibinfo {pages} {100402} (\bibinfo {year} {2020})}\BibitemShut
  {NoStop}%
\bibitem [{\citenamefont
  {Haine}(2013)}]{haineInformationRecyclingBeamSplitters2013}%
  \BibitemOpen
  \bibfield  {author} {\bibinfo {author} {\bibfnamefont {S.~A.}\ \bibnamefont
  {Haine}},\ }\bibfield  {title} {\bibinfo {title} {Information-{{Recycling
  Beam Splitters}} for {{Quantum Enhanced Atom Interferometry}}},\ }\href
  {https://doi.org/10.1103/PhysRevLett.110.053002} {\bibfield  {journal}
  {\bibinfo  {journal} {Physical Review Letters}\ }\textbf {\bibinfo {volume}
  {110}},\ \bibinfo {pages} {053002} (\bibinfo {year} {2013})}\BibitemShut
  {NoStop}%
\bibitem [{\citenamefont {Haine}\ and\ \citenamefont
  {Lau}(2016)}]{haineGenerationAtomlightEntanglement2016}%
  \BibitemOpen
  \bibfield  {author} {\bibinfo {author} {\bibfnamefont {S.~A.}\ \bibnamefont
  {Haine}}\ and\ \bibinfo {author} {\bibfnamefont {W.~Y.~S.}\ \bibnamefont
  {Lau}},\ }\bibfield  {title} {\bibinfo {title} {Generation of atom-light
  entanglement in an optical cavity for quantum enhanced atom interferometry},\
  }\href {https://doi.org/10.1103/PhysRevA.93.023607} {\bibfield  {journal}
  {\bibinfo  {journal} {Physical Review A}\ }\textbf {\bibinfo {volume} {93}},\
  \bibinfo {pages} {023607} (\bibinfo {year} {2016})}\BibitemShut {NoStop}%
\bibitem [{\citenamefont {Kritsotakis}\ \emph {et~al.}(2021)\citenamefont
  {Kritsotakis}, \citenamefont {Dunningham},\ and\ \citenamefont
  {Haine}}]{kritsotakisSpinSqueezingBoseEinstein2021}%
  \BibitemOpen
  \bibfield  {author} {\bibinfo {author} {\bibfnamefont {M.}~\bibnamefont
  {Kritsotakis}}, \bibinfo {author} {\bibfnamefont {J.~A.}\ \bibnamefont
  {Dunningham}},\ and\ \bibinfo {author} {\bibfnamefont {S.~A.}\ \bibnamefont
  {Haine}},\ }\bibfield  {title} {\bibinfo {title} {Spin squeezing of a
  {{Bose-Einstein}} condensate via a quantum nondemolition measurement for
  quantum-enhanced atom interferometry},\ }\href
  {https://doi.org/10.1103/PhysRevA.103.023318} {\bibfield  {journal} {\bibinfo
   {journal} {Physical Review A}\ }\textbf {\bibinfo {volume} {103}},\ \bibinfo
  {pages} {023318} (\bibinfo {year} {2021})}\BibitemShut {NoStop}%
\bibitem [{\citenamefont {Fuderer}\ \emph {et~al.}(2023)\citenamefont
  {Fuderer}, \citenamefont {Hope},\ and\ \citenamefont
  {Haine}}]{fudererHybridMethodGenerating2023a}%
  \BibitemOpen
  \bibfield  {author} {\bibinfo {author} {\bibfnamefont {L.~A.}\ \bibnamefont
  {Fuderer}}, \bibinfo {author} {\bibfnamefont {J.~J.}\ \bibnamefont {Hope}},\
  and\ \bibinfo {author} {\bibfnamefont {S.~A.}\ \bibnamefont {Haine}},\
  }\bibfield  {title} {\bibinfo {title} {Hybrid method of generating
  spin-squeezed states for quantum-enhanced atom interferometry},\ }\href
  {https://doi.org/10.1103/PhysRevA.108.043722} {\bibfield  {journal} {\bibinfo
   {journal} {Physical Review A}\ }\textbf {\bibinfo {volume} {108}},\ \bibinfo
  {pages} {043722} (\bibinfo {year} {2023})}\BibitemShut {NoStop}%
\bibitem [{\citenamefont {Blakie}\ and\ \citenamefont
  {Davis}(2005)}]{blakieProjectedGrossPitaevskiiEquation2005}%
  \BibitemOpen
  \bibfield  {author} {\bibinfo {author} {\bibfnamefont {P.~B.}\ \bibnamefont
  {Blakie}}\ and\ \bibinfo {author} {\bibfnamefont {M.~J.}\ \bibnamefont
  {Davis}},\ }\bibfield  {title} {\bibinfo {title} {Projected
  {{Gross-Pitaevskii}} equation for harmonically confined {{Bose}} gases at
  finite temperature},\ }\href {https://doi.org/10.1103/PhysRevA.72.063608}
  {\bibfield  {journal} {\bibinfo  {journal} {Physical Review A}\ }\textbf
  {\bibinfo {volume} {72}},\ \bibinfo {pages} {063608} (\bibinfo {year}
  {2005})}\BibitemShut {NoStop}%
\bibitem [{\citenamefont {Weiler}\ \emph {et~al.}(2008)\citenamefont {Weiler},
  \citenamefont {Neely}, \citenamefont {Scherer}, \citenamefont {Bradley},
  \citenamefont {Davis},\ and\ \citenamefont
  {Anderson}}]{weilerSpontaneousVorticesFormation2008}%
  \BibitemOpen
  \bibfield  {author} {\bibinfo {author} {\bibfnamefont {C.~N.}\ \bibnamefont
  {Weiler}}, \bibinfo {author} {\bibfnamefont {T.~W.}\ \bibnamefont {Neely}},
  \bibinfo {author} {\bibfnamefont {D.~R.}\ \bibnamefont {Scherer}}, \bibinfo
  {author} {\bibfnamefont {A.~S.}\ \bibnamefont {Bradley}}, \bibinfo {author}
  {\bibfnamefont {M.~J.}\ \bibnamefont {Davis}},\ and\ \bibinfo {author}
  {\bibfnamefont {B.~P.}\ \bibnamefont {Anderson}},\ }\bibfield  {title}
  {\bibinfo {title} {Spontaneous vortices in the formation of
  {{Bose}}--{{Einstein}} condensates},\ }\href
  {https://doi.org/10.1038/nature07334} {\bibfield  {journal} {\bibinfo
  {journal} {Nature}\ }\textbf {\bibinfo {volume} {455}},\ \bibinfo {pages}
  {948} (\bibinfo {year} {2008})}\BibitemShut {NoStop}%
\bibitem [{\citenamefont {Rooney}\ \emph {et~al.}(2012)\citenamefont {Rooney},
  \citenamefont {Blakie},\ and\ \citenamefont
  {Bradley}}]{rooneyStochasticProjectedGrossPitaevskii2012}%
  \BibitemOpen
  \bibfield  {author} {\bibinfo {author} {\bibfnamefont {S.~J.}\ \bibnamefont
  {Rooney}}, \bibinfo {author} {\bibfnamefont {P.~B.}\ \bibnamefont {Blakie}},\
  and\ \bibinfo {author} {\bibfnamefont {A.~S.}\ \bibnamefont {Bradley}},\
  }\bibfield  {title} {\bibinfo {title} {Stochastic projected
  {{Gross-Pitaevskii}} equation},\ }\href
  {https://doi.org/10.1103/PhysRevA.86.053634} {\bibfield  {journal} {\bibinfo
  {journal} {Physical Review A}\ }\textbf {\bibinfo {volume} {86}},\ \bibinfo
  {pages} {053634} (\bibinfo {year} {2012})}\BibitemShut {NoStop}%
\bibitem [{\citenamefont {Hush}(2012)}]{hushEfficientSimulationControlled2012}%
  \BibitemOpen
  \bibfield  {author} {\bibinfo {author} {\bibfnamefont {M.~R.}\ \bibnamefont
  {Hush}},\ }\bibfield  {title} {\bibinfo {title} {Efficient simulation of
  controlled large quantum systems}\ }\href
  {https://doi.org/10.25911/5d514d3164727} {10.25911/5d514d3164727} (\bibinfo
  {year} {2012})\BibitemShut {NoStop}%
\bibitem [{\citenamefont {Jacobs}\ \emph {et~al.}(2014)\citenamefont {Jacobs},
  \citenamefont {Wang},\ and\ \citenamefont
  {Wiseman}}]{jacobsCoherentFeedbackThat2014}%
  \BibitemOpen
  \bibfield  {author} {\bibinfo {author} {\bibfnamefont {K.}~\bibnamefont
  {Jacobs}}, \bibinfo {author} {\bibfnamefont {X.}~\bibnamefont {Wang}},\ and\
  \bibinfo {author} {\bibfnamefont {H.~M.}\ \bibnamefont {Wiseman}},\
  }\bibfield  {title} {\bibinfo {title} {Coherent feedback that beats all
  measurement-based feedback protocols},\ }\href
  {https://doi.org/10.1088/1367-2630/16/7/073036} {\bibfield  {journal}
  {\bibinfo  {journal} {New Journal of Physics}\ }\textbf {\bibinfo {volume}
  {16}},\ \bibinfo {pages} {073036} (\bibinfo {year} {2014})}\BibitemShut
  {NoStop}%
\bibitem [{\citenamefont {Lloyd}(2000)}]{lloydCoherentQuantumFeedback2000}%
  \BibitemOpen
  \bibfield  {author} {\bibinfo {author} {\bibfnamefont {S.}~\bibnamefont
  {Lloyd}},\ }\bibfield  {title} {\bibinfo {title} {Coherent quantum
  feedback},\ }\href {https://doi.org/10.1103/PhysRevA.62.022108} {\bibfield
  {journal} {\bibinfo  {journal} {Physical Review A}\ }\textbf {\bibinfo
  {volume} {62}},\ \bibinfo {pages} {022108} (\bibinfo {year}
  {2000})}\BibitemShut {NoStop}%
\bibitem [{\citenamefont {Gati}\ and\ \citenamefont
  {Oberthaler}(2007)}]{gatiBosonicJosephsonJunction2007}%
  \BibitemOpen
  \bibfield  {author} {\bibinfo {author} {\bibfnamefont {R.}~\bibnamefont
  {Gati}}\ and\ \bibinfo {author} {\bibfnamefont {M.~K.}\ \bibnamefont
  {Oberthaler}},\ }\bibfield  {title} {\bibinfo {title} {A bosonic
  {{Josephson}} junction},\ }\href
  {https://doi.org/10.1088/0953-4075/40/10/R01} {\bibfield  {journal} {\bibinfo
   {journal} {Journal of Physics B: Atomic, Molecular and Optical Physics}\
  }\textbf {\bibinfo {volume} {40}},\ \bibinfo {pages} {R61} (\bibinfo {year}
  {2007})}\BibitemShut {NoStop}%
\bibitem [{\citenamefont {Milburn}\ \emph {et~al.}(1997)\citenamefont
  {Milburn}, \citenamefont {Corney}, \citenamefont {Wright},\ and\
  \citenamefont {Walls}}]{milburnQuantumDynamicsAtomic1997}%
  \BibitemOpen
  \bibfield  {author} {\bibinfo {author} {\bibfnamefont {G.~J.}\ \bibnamefont
  {Milburn}}, \bibinfo {author} {\bibfnamefont {J.}~\bibnamefont {Corney}},
  \bibinfo {author} {\bibfnamefont {E.~M.}\ \bibnamefont {Wright}},\ and\
  \bibinfo {author} {\bibfnamefont {D.~F.}\ \bibnamefont {Walls}},\ }\bibfield
  {title} {\bibinfo {title} {Quantum dynamics of an atomic {{Bose-Einstein}}
  condensate in a double-well potential},\ }\href
  {https://doi.org/10.1103/PhysRevA.55.4318} {\bibfield  {journal} {\bibinfo
  {journal} {Physical Review A}\ }\textbf {\bibinfo {volume} {55}},\ \bibinfo
  {pages} {4318} (\bibinfo {year} {1997})}\BibitemShut {NoStop}%
\bibitem [{\citenamefont {Radzihovsky}\ and\ \citenamefont
  {Gurarie}(2010)}]{radzihovskyRelationAcJosephson2010}%
  \BibitemOpen
  \bibfield  {author} {\bibinfo {author} {\bibfnamefont {L.}~\bibnamefont
  {Radzihovsky}}\ and\ \bibinfo {author} {\bibfnamefont {V.}~\bibnamefont
  {Gurarie}},\ }\bibfield  {title} {\bibinfo {title} {Relation between ac
  {{Josephson}} effect and double-well {{Bose-Einstein-condensate}}
  oscillations},\ }\href {https://doi.org/10.1103/PhysRevA.81.063609}
  {\bibfield  {journal} {\bibinfo  {journal} {Physical Review A}\ }\textbf
  {\bibinfo {volume} {81}},\ \bibinfo {pages} {063609} (\bibinfo {year}
  {2010})}\BibitemShut {NoStop}%
\bibitem [{\citenamefont {Borah}\ \emph {et~al.}(2021)\citenamefont {Borah},
  \citenamefont {Sarma}, \citenamefont {Kewming}, \citenamefont {Milburn},\
  and\ \citenamefont {Twamley}}]{borahMeasurementBasedFeedbackQuantum2021}%
  \BibitemOpen
  \bibfield  {author} {\bibinfo {author} {\bibfnamefont {S.}~\bibnamefont
  {Borah}}, \bibinfo {author} {\bibfnamefont {B.}~\bibnamefont {Sarma}},
  \bibinfo {author} {\bibfnamefont {M.}~\bibnamefont {Kewming}}, \bibinfo
  {author} {\bibfnamefont {G.~J.}\ \bibnamefont {Milburn}},\ and\ \bibinfo
  {author} {\bibfnamefont {J.}~\bibnamefont {Twamley}},\ }\bibfield  {title}
  {\bibinfo {title} {Measurement-{{Based Feedback Quantum Control}} with {{Deep
  Reinforcement Learning}} for a {{Double-Well Nonlinear Potential}}},\ }\href
  {https://doi.org/10.1103/PhysRevLett.127.190403} {\bibfield  {journal}
  {\bibinfo  {journal} {Physical Review Letters}\ }\textbf {\bibinfo {volume}
  {127}},\ \bibinfo {pages} {190403} (\bibinfo {year} {2021})}\BibitemShut
  {NoStop}%
\bibitem [{\citenamefont {Borah}\ and\ \citenamefont
  {Sarma}(2023)}]{borahNoCollapseAccurateQuantum2023}%
  \BibitemOpen
  \bibfield  {author} {\bibinfo {author} {\bibfnamefont {S.}~\bibnamefont
  {Borah}}\ and\ \bibinfo {author} {\bibfnamefont {B.}~\bibnamefont {Sarma}},\
  }\bibfield  {title} {\bibinfo {title} {No-{{Collapse Accurate Quantum
  Feedback Control}} via {{Conditional State Tomography}}},\ }\href
  {https://doi.org/10.1103/PhysRevLett.131.210803} {\bibfield  {journal}
  {\bibinfo  {journal} {Physical Review Letters}\ }\textbf {\bibinfo {volume}
  {131}},\ \bibinfo {pages} {210803} (\bibinfo {year} {2023})}\BibitemShut
  {NoStop}%
\bibitem [{\citenamefont {Gross}(2012)}]{grossSpinSqueezingEntanglement2012}%
  \BibitemOpen
  \bibfield  {author} {\bibinfo {author} {\bibfnamefont {C.}~\bibnamefont
  {Gross}},\ }\bibfield  {title} {\bibinfo {title} {Spin squeezing,
  entanglement and quantum metrology with {{Bose}}--{{Einstein}} condensates},\
  }\href {https://doi.org/10.1088/0953-4075/45/10/103001} {\bibfield  {journal}
  {\bibinfo  {journal} {Journal of Physics B: Atomic, Molecular and Optical
  Physics}\ }\textbf {\bibinfo {volume} {45}},\ \bibinfo {pages} {103001}
  (\bibinfo {year} {2012})}\BibitemShut {NoStop}%
\bibitem [{\citenamefont {Arecchi}\ \emph {et~al.}(1972)\citenamefont
  {Arecchi}, \citenamefont {Courtens}, \citenamefont {Gilmore},\ and\
  \citenamefont {Thomas}}]{arecchiAtomicCoherentStates1972}%
  \BibitemOpen
  \bibfield  {author} {\bibinfo {author} {\bibfnamefont {F.~T.}\ \bibnamefont
  {Arecchi}}, \bibinfo {author} {\bibfnamefont {E.}~\bibnamefont {Courtens}},
  \bibinfo {author} {\bibfnamefont {R.}~\bibnamefont {Gilmore}},\ and\ \bibinfo
  {author} {\bibfnamefont {H.}~\bibnamefont {Thomas}},\ }\bibfield  {title}
  {\bibinfo {title} {Atomic {{Coherent States}} in {{Quantum Optics}}},\ }\href
  {https://doi.org/10.1103/PhysRevA.6.2211} {\bibfield  {journal} {\bibinfo
  {journal} {Physical Review A}\ }\textbf {\bibinfo {volume} {6}},\ \bibinfo
  {pages} {2211} (\bibinfo {year} {1972})}\BibitemShut {NoStop}%
\bibitem [{\citenamefont {Hush}\ \emph {et~al.}(2010)\citenamefont {Hush},
  \citenamefont {Carvalho},\ and\ \citenamefont
  {Hope}}]{hushNumberphaseWignerRepresentation2010}%
  \BibitemOpen
  \bibfield  {author} {\bibinfo {author} {\bibfnamefont {M.~R.}\ \bibnamefont
  {Hush}}, \bibinfo {author} {\bibfnamefont {A.~R.~R.}\ \bibnamefont
  {Carvalho}},\ and\ \bibinfo {author} {\bibfnamefont {J.~J.}\ \bibnamefont
  {Hope}},\ }\bibfield  {title} {\bibinfo {title} {Number-phase {{Wigner}}
  representation for efficient stochastic simulations},\ }\href
  {https://doi.org/10.1103/PhysRevA.81.033852} {\bibfield  {journal} {\bibinfo
  {journal} {Physical Review A}\ }\textbf {\bibinfo {volume} {81}},\ \bibinfo
  {pages} {033852} (\bibinfo {year} {2010})}\BibitemShut {NoStop}%
\bibitem [{\citenamefont {{Ilo-Okeke}}\ and\ \citenamefont
  {Byrnes}(2014)}]{ilo-okekeTheorySingleShotPhase2014}%
  \BibitemOpen
  \bibfield  {author} {\bibinfo {author} {\bibfnamefont {E.~O.}\ \bibnamefont
  {{Ilo-Okeke}}}\ and\ \bibinfo {author} {\bibfnamefont {T.}~\bibnamefont
  {Byrnes}},\ }\bibfield  {title} {\bibinfo {title} {Theory of {{Single-Shot
  Phase Contrast Imaging}} in {{Spinor Bose-Einstein Condensates}}},\ }\href
  {https://doi.org/10.1103/PhysRevLett.112.233602} {\bibfield  {journal}
  {\bibinfo  {journal} {Physical Review Letters}\ }\textbf {\bibinfo {volume}
  {112}},\ \bibinfo {pages} {233602} (\bibinfo {year} {2014})}\BibitemShut
  {NoStop}%
\bibitem [{\citenamefont {Penrose}\ and\ \citenamefont
  {Onsager}(1956)}]{penroseBoseEinsteinCondensationLiquid1956}%
  \BibitemOpen
  \bibfield  {author} {\bibinfo {author} {\bibfnamefont {O.}~\bibnamefont
  {Penrose}}\ and\ \bibinfo {author} {\bibfnamefont {L.}~\bibnamefont
  {Onsager}},\ }\bibfield  {title} {\bibinfo {title} {Bose-{{Einstein
  Condensation}} and {{Liquid Helium}}},\ }\href
  {https://doi.org/10.1103/PhysRev.104.576} {\bibfield  {journal} {\bibinfo
  {journal} {Physical Review}\ }\textbf {\bibinfo {volume} {104}},\ \bibinfo
  {pages} {576} (\bibinfo {year} {1956})}\BibitemShut {NoStop}%
\bibitem [{\citenamefont {Gauthier}\ \emph {et~al.}(2016)\citenamefont
  {Gauthier}, \citenamefont {Lenton}, \citenamefont {Parry}, \citenamefont
  {Baker}, \citenamefont {Davis}, \citenamefont {{Rubinsztein-Dunlop}},\ and\
  \citenamefont {Neely}}]{gauthierDirectImagingDigitalmicromirror2016}%
  \BibitemOpen
  \bibfield  {author} {\bibinfo {author} {\bibfnamefont {G.}~\bibnamefont
  {Gauthier}}, \bibinfo {author} {\bibfnamefont {I.}~\bibnamefont {Lenton}},
  \bibinfo {author} {\bibfnamefont {N.~M.}\ \bibnamefont {Parry}}, \bibinfo
  {author} {\bibfnamefont {M.}~\bibnamefont {Baker}}, \bibinfo {author}
  {\bibfnamefont {M.~J.}\ \bibnamefont {Davis}}, \bibinfo {author}
  {\bibfnamefont {H.}~\bibnamefont {{Rubinsztein-Dunlop}}},\ and\ \bibinfo
  {author} {\bibfnamefont {T.~W.}\ \bibnamefont {Neely}},\ }\bibfield  {title}
  {\bibinfo {title} {Direct imaging of a digital-micromirror device for
  configurable microscopic optical potentials},\ }\href
  {https://doi.org/10.1364/OPTICA.3.001136} {\bibfield  {journal} {\bibinfo
  {journal} {Optica}\ }\textbf {\bibinfo {volume} {3}},\ \bibinfo {pages}
  {1136} (\bibinfo {year} {2016})}\BibitemShut {NoStop}%
\bibitem [{\citenamefont {Gauthier}\ \emph {et~al.}(2021)\citenamefont
  {Gauthier}, \citenamefont {Bell}, \citenamefont {Stilgoe}, \citenamefont
  {Baker}, \citenamefont {{Rubinsztein-Dunlop}},\ and\ \citenamefont
  {Neely}}]{gauthierDynamicHighresolutionOptical2021}%
  \BibitemOpen
  \bibfield  {author} {\bibinfo {author} {\bibfnamefont {G.}~\bibnamefont
  {Gauthier}}, \bibinfo {author} {\bibfnamefont {T.~A.}\ \bibnamefont {Bell}},
  \bibinfo {author} {\bibfnamefont {A.~B.}\ \bibnamefont {Stilgoe}}, \bibinfo
  {author} {\bibfnamefont {M.}~\bibnamefont {Baker}}, \bibinfo {author}
  {\bibfnamefont {H.}~\bibnamefont {{Rubinsztein-Dunlop}}},\ and\ \bibinfo
  {author} {\bibfnamefont {T.~W.}\ \bibnamefont {Neely}},\ }\bibfield  {title}
  {\bibinfo {title} {Dynamic high-resolution optical trapping of ultracold
  atoms},\ }\href {https://doi.org/10.1016/bs.aamop.2021.04.001} {\bibfield
  {journal} {\bibinfo  {journal} {Advances in Atomic, Molecular and Optical
  Physics}\ }\textbf {\bibinfo {volume} {70}},\ \bibinfo {pages} {1} (\bibinfo
  {year} {2021})}\BibitemShut {NoStop}%
\bibitem [{\citenamefont
  {Szigeti}(2013)}]{szigetiControlledBosecondensedSources2013}%
  \BibitemOpen
  \bibfield  {author} {\bibinfo {author} {\bibfnamefont {S.~S.}\ \bibnamefont
  {Szigeti}},\ }\bibfield  {title} {\bibinfo {title} {Controlled
  {{Bose-condensed}} sources for atom interferometry}\ }\href
  {https://doi.org/10.25911/5d514a57ce679} {10.25911/5d514a57ce679} (\bibinfo
  {year} {2013})\BibitemShut {NoStop}%
\bibitem [{\citenamefont {Pethick}\ and\ \citenamefont
  {Smith}(2008)}]{pethickBoseEinsteinCondensation2008}%
  \BibitemOpen
  \bibfield  {author} {\bibinfo {author} {\bibfnamefont {C.~J.}\ \bibnamefont
  {Pethick}}\ and\ \bibinfo {author} {\bibfnamefont {H.}~\bibnamefont
  {Smith}},\ }\href@noop {} {\emph {\bibinfo {title} {Bose--{{Einstein
  Condensation}} in {{Dilute Gases}}}}},\ \bibinfo {edition} {2nd}\ ed.\
  (\bibinfo  {publisher} {Cambridge University Press},\ \bibinfo {address}
  {Cambridge ; New York},\ \bibinfo {year} {2008})\BibitemShut {NoStop}%
\bibitem [{\citenamefont
  {Mehdi}(2024)}]{mehdiSuperfluidDissipationFeedback2024}%
  \BibitemOpen
  \bibfield  {author} {\bibinfo {author} {\bibfnamefont {Z.}~\bibnamefont
  {Mehdi}},\ }\bibfield  {title} {\bibinfo {title} {Superfluid {{Dissipation}}
  and {{Feedback Cooling}} in {{Ultracold Atomic Gases}}}\ }\href
  {https://doi.org/10.25911/XFRG-QV06} {10.25911/XFRG-QV06} (\bibinfo {year}
  {2024})\BibitemShut {NoStop}%
\bibitem [{\citenamefont {Zhu}(2023)}]{zhuFeedbackCoolingDegenerate2023}%
  \BibitemOpen
  \bibfield  {author} {\bibinfo {author} {\bibfnamefont {K.}~\bibnamefont
  {Zhu}},\ }\bibfield  {title} {\bibinfo {title} {Feedback {{Cooling}} of
  {{Degenerate Quantum Gases}}}\ }\href {https://doi.org/10.25911/254M-G795}
  {10.25911/254M-G795} (\bibinfo {year} {2023})\BibitemShut {NoStop}%
\bibitem [{\citenamefont {Rooney}\ \emph {et~al.}(2014)\citenamefont {Rooney},
  \citenamefont {Blakie},\ and\ \citenamefont
  {Bradley}}]{rooneyNumericalMethodStochastic2014}%
  \BibitemOpen
  \bibfield  {author} {\bibinfo {author} {\bibfnamefont {S.~J.}\ \bibnamefont
  {Rooney}}, \bibinfo {author} {\bibfnamefont {P.~B.}\ \bibnamefont {Blakie}},\
  and\ \bibinfo {author} {\bibfnamefont {A.~S.}\ \bibnamefont {Bradley}},\
  }\bibfield  {title} {\bibinfo {title} {Numerical method for the stochastic
  projected {{Gross-Pitaevskii}} equation},\ }\href
  {https://doi.org/10.1103/PhysRevE.89.013302} {\bibfield  {journal} {\bibinfo
  {journal} {Physical Review E}\ }\textbf {\bibinfo {volume} {89}},\ \bibinfo
  {pages} {013302} (\bibinfo {year} {2014})}\BibitemShut {NoStop}%
\bibitem [{\citenamefont {Bradley}\ \emph {et~al.}(2015)\citenamefont
  {Bradley}, \citenamefont {Rooney},\ and\ \citenamefont
  {McDonald}}]{bradleyLowdimensionalStochasticProjected2015}%
  \BibitemOpen
  \bibfield  {author} {\bibinfo {author} {\bibfnamefont {A.~S.}\ \bibnamefont
  {Bradley}}, \bibinfo {author} {\bibfnamefont {S.~J.}\ \bibnamefont
  {Rooney}},\ and\ \bibinfo {author} {\bibfnamefont {R.~G.}\ \bibnamefont
  {McDonald}},\ }\bibfield  {title} {\bibinfo {title} {Low-dimensional
  stochastic projected {{Gross-Pitaevskii}} equation},\ }\href
  {https://doi.org/10.1103/PhysRevA.92.033631} {\bibfield  {journal} {\bibinfo
  {journal} {Physical Review A}\ }\textbf {\bibinfo {volume} {92}},\ \bibinfo
  {pages} {033631} (\bibinfo {year} {2015})}\BibitemShut {NoStop}%
\bibitem [{\citenamefont {Pitaevskii}\ \emph {et~al.}(2003)\citenamefont
  {Pitaevskii}, \citenamefont {Stringari}, \citenamefont {Pitaevskii},\ and\
  \citenamefont {Stringari}}]{pitaevskiiBoseEinsteinCondensation2003}%
  \BibitemOpen
  \bibfield  {author} {\bibinfo {author} {\bibfnamefont {L.~P.}\ \bibnamefont
  {Pitaevskii}}, \bibinfo {author} {\bibfnamefont {S.}~\bibnamefont
  {Stringari}}, \bibinfo {author} {\bibfnamefont {L.~P.}\ \bibnamefont
  {Pitaevskii}},\ and\ \bibinfo {author} {\bibfnamefont {S.}~\bibnamefont
  {Stringari}},\ }\href@noop {} {\emph {\bibinfo {title} {Bose-{{Einstein
  Condensation}}}}},\ International {{Series}} of {{Monographs}} on
  {{Physics}}\ (\bibinfo  {publisher} {Oxford University Press},\ \bibinfo
  {address} {Oxford, New York},\ \bibinfo {year} {2003})\BibitemShut {NoStop}%
\bibitem [{\citenamefont {Aspelmeyer}\ \emph {et~al.}(2014)\citenamefont
  {Aspelmeyer}, \citenamefont {Kippenberg},\ and\ \citenamefont
  {Marquardt}}]{aspelmeyerCavityOptomechanics2014}%
  \BibitemOpen
  \bibfield  {author} {\bibinfo {author} {\bibfnamefont {M.}~\bibnamefont
  {Aspelmeyer}}, \bibinfo {author} {\bibfnamefont {T.~J.}\ \bibnamefont
  {Kippenberg}},\ and\ \bibinfo {author} {\bibfnamefont {F.}~\bibnamefont
  {Marquardt}},\ }\bibfield  {title} {\bibinfo {title} {Cavity optomechanics},\
  }\href {https://doi.org/10.1103/RevModPhys.86.1391} {\bibfield  {journal}
  {\bibinfo  {journal} {Reviews of Modern Physics}\ }\textbf {\bibinfo {volume}
  {86}},\ \bibinfo {pages} {1391} (\bibinfo {year} {2014})}\BibitemShut
  {NoStop}%
\bibitem [{\citenamefont {Poyatos}\ \emph {et~al.}(1996)\citenamefont
  {Poyatos}, \citenamefont {Cirac},\ and\ \citenamefont
  {Zoller}}]{poyatosQuantumReservoirEngineering1996}%
  \BibitemOpen
  \bibfield  {author} {\bibinfo {author} {\bibfnamefont {J.~F.}\ \bibnamefont
  {Poyatos}}, \bibinfo {author} {\bibfnamefont {J.~I.}\ \bibnamefont {Cirac}},\
  and\ \bibinfo {author} {\bibfnamefont {P.}~\bibnamefont {Zoller}},\
  }\bibfield  {title} {\bibinfo {title} {Quantum {{Reservoir Engineering}} with
  {{Laser Cooled Trapped Ions}}},\ }\href
  {https://doi.org/10.1103/PhysRevLett.77.4728} {\bibfield  {journal} {\bibinfo
   {journal} {Physical Review Letters}\ }\textbf {\bibinfo {volume} {77}},\
  \bibinfo {pages} {4728} (\bibinfo {year} {1996})}\BibitemShut {NoStop}%
\bibitem [{\citenamefont {Basilewitsch}\ \emph {et~al.}(2019)\citenamefont
  {Basilewitsch}, \citenamefont {Cosco}, \citenamefont {Gullo}, \citenamefont
  {M{\"o}tt{\"o}nen}, \citenamefont {{Ala-Nissil{\"a}}}, \citenamefont {Koch},\
  and\ \citenamefont {Maniscalco}}]{basilewitschReservoirEngineeringUsing2019}%
  \BibitemOpen
  \bibfield  {author} {\bibinfo {author} {\bibfnamefont {D.}~\bibnamefont
  {Basilewitsch}}, \bibinfo {author} {\bibfnamefont {F.}~\bibnamefont {Cosco}},
  \bibinfo {author} {\bibfnamefont {N.~L.}\ \bibnamefont {Gullo}}, \bibinfo
  {author} {\bibfnamefont {M.}~\bibnamefont {M{\"o}tt{\"o}nen}}, \bibinfo
  {author} {\bibfnamefont {T.}~\bibnamefont {{Ala-Nissil{\"a}}}}, \bibinfo
  {author} {\bibfnamefont {C.~P.}\ \bibnamefont {Koch}},\ and\ \bibinfo
  {author} {\bibfnamefont {S.}~\bibnamefont {Maniscalco}},\ }\bibfield  {title}
  {\bibinfo {title} {Reservoir engineering using quantum optimal control for
  qubit reset},\ }\href {https://doi.org/10.1088/1367-2630/ab41ad} {\bibfield
  {journal} {\bibinfo  {journal} {New Journal of Physics}\ }\textbf {\bibinfo
  {volume} {21}},\ \bibinfo {pages} {093054} (\bibinfo {year}
  {2019})}\BibitemShut {NoStop}%
\bibitem [{\citenamefont {Tissot}\ \emph {et~al.}(2024)\citenamefont {Tissot},
  \citenamefont {Ribeiro},\ and\ \citenamefont
  {Marquardt}}]{tissotReservoirEngineeringClassical2024}%
  \BibitemOpen
  \bibfield  {author} {\bibinfo {author} {\bibfnamefont {B.}~\bibnamefont
  {Tissot}}, \bibinfo {author} {\bibfnamefont {H.}~\bibnamefont {Ribeiro}},\
  and\ \bibinfo {author} {\bibfnamefont {F.}~\bibnamefont {Marquardt}},\
  }\bibfield  {title} {\bibinfo {title} {Reservoir engineering for classical
  nonlinear fields},\ }\href {https://doi.org/10.1103/PhysRevResearch.6.023015}
  {\bibfield  {journal} {\bibinfo  {journal} {Physical Review Research}\
  }\textbf {\bibinfo {volume} {6}},\ \bibinfo {pages} {023015} (\bibinfo {year}
  {2024})}\BibitemShut {NoStop}%
\bibitem [{\citenamefont {van Handel}\ \emph {et~al.}(2005)\citenamefont {van
  Handel}, \citenamefont {Stockton},\ and\ \citenamefont
  {Mabuchi}}]{handelModellingFeedbackControl2005}%
  \BibitemOpen
  \bibfield  {author} {\bibinfo {author} {\bibfnamefont {R.}~\bibnamefont {van
  Handel}}, \bibinfo {author} {\bibfnamefont {J.~K.}\ \bibnamefont
  {Stockton}},\ and\ \bibinfo {author} {\bibfnamefont {H.}~\bibnamefont
  {Mabuchi}},\ }\bibfield  {title} {\bibinfo {title} {Modelling and feedback
  control design for quantum state preparation},\ }\href
  {https://doi.org/10.1088/1464-4266/7/10/001} {\bibfield  {journal} {\bibinfo
  {journal} {Journal of Optics B: Quantum and Semiclassical Optics}\ }\textbf
  {\bibinfo {volume} {7}},\ \bibinfo {pages} {S179} (\bibinfo {year}
  {2005})}\BibitemShut {NoStop}%
\bibitem [{\citenamefont {Mills}\ \emph {et~al.}(2022)\citenamefont {Mills},
  \citenamefont {Guinn}, \citenamefont {Feldman}, \citenamefont {Sigillito},
  \citenamefont {Gullans}, \citenamefont {Rakher}, \citenamefont {Kerckhoff},
  \citenamefont {Jackson},\ and\ \citenamefont
  {Petta}}]{millsHighFidelityStatePreparation2022}%
  \BibitemOpen
  \bibfield  {author} {\bibinfo {author} {\bibfnamefont {A.}~\bibnamefont
  {Mills}}, \bibinfo {author} {\bibfnamefont {C.}~\bibnamefont {Guinn}},
  \bibinfo {author} {\bibfnamefont {M.}~\bibnamefont {Feldman}}, \bibinfo
  {author} {\bibfnamefont {A.}~\bibnamefont {Sigillito}}, \bibinfo {author}
  {\bibfnamefont {M.}~\bibnamefont {Gullans}}, \bibinfo {author} {\bibfnamefont
  {M.}~\bibnamefont {Rakher}}, \bibinfo {author} {\bibfnamefont
  {J.}~\bibnamefont {Kerckhoff}}, \bibinfo {author} {\bibfnamefont
  {C.}~\bibnamefont {Jackson}},\ and\ \bibinfo {author} {\bibfnamefont
  {J.}~\bibnamefont {Petta}},\ }\bibfield  {title} {\bibinfo {title}
  {High-{{Fidelity State Preparation}}, {{Quantum Control}}, and {{Readout}} of
  an {{Isotopically Enriched Silicon Spin Qubit}}},\ }\href
  {https://doi.org/10.1103/PhysRevApplied.18.064028} {\bibfield  {journal}
  {\bibinfo  {journal} {Physical Review Applied}\ }\textbf {\bibinfo {volume}
  {18}},\ \bibinfo {pages} {064028} (\bibinfo {year} {2022})}\BibitemShut
  {NoStop}%
\bibitem [{\citenamefont {Sayrin}\ \emph {et~al.}(2011)\citenamefont {Sayrin},
  \citenamefont {Dotsenko}, \citenamefont {Zhou}, \citenamefont {Peaudecerf},
  \citenamefont {Rybarczyk}, \citenamefont {Gleyzes}, \citenamefont {Rouchon},
  \citenamefont {Mirrahimi}, \citenamefont {Amini}, \citenamefont {Brune},
  \citenamefont {Raimond},\ and\ \citenamefont
  {Haroche}}]{sayrinRealtimeQuantumFeedback2011}%
  \BibitemOpen
  \bibfield  {author} {\bibinfo {author} {\bibfnamefont {C.}~\bibnamefont
  {Sayrin}}, \bibinfo {author} {\bibfnamefont {I.}~\bibnamefont {Dotsenko}},
  \bibinfo {author} {\bibfnamefont {X.}~\bibnamefont {Zhou}}, \bibinfo {author}
  {\bibfnamefont {B.}~\bibnamefont {Peaudecerf}}, \bibinfo {author}
  {\bibfnamefont {T.}~\bibnamefont {Rybarczyk}}, \bibinfo {author}
  {\bibfnamefont {S.}~\bibnamefont {Gleyzes}}, \bibinfo {author} {\bibfnamefont
  {P.}~\bibnamefont {Rouchon}}, \bibinfo {author} {\bibfnamefont
  {M.}~\bibnamefont {Mirrahimi}}, \bibinfo {author} {\bibfnamefont
  {H.}~\bibnamefont {Amini}}, \bibinfo {author} {\bibfnamefont
  {M.}~\bibnamefont {Brune}}, \bibinfo {author} {\bibfnamefont {J.-M.}\
  \bibnamefont {Raimond}},\ and\ \bibinfo {author} {\bibfnamefont
  {S.}~\bibnamefont {Haroche}},\ }\bibfield  {title} {\bibinfo {title}
  {Real-time quantum feedback prepares and stabilizes photon number states},\
  }\href {https://doi.org/10.1038/nature10376} {\bibfield  {journal} {\bibinfo
  {journal} {Nature}\ }\textbf {\bibinfo {volume} {477}},\ \bibinfo {pages}
  {73} (\bibinfo {year} {2011})}\BibitemShut {NoStop}%
\bibitem [{\citenamefont {Guo}\ \emph {et~al.}(2023)\citenamefont {Guo},
  \citenamefont {Chang}, \citenamefont {Yao},\ and\ \citenamefont
  {Gr{\"o}blacher}}]{guoActivefeedbackQuantumControl2023}%
  \BibitemOpen
  \bibfield  {author} {\bibinfo {author} {\bibfnamefont {J.}~\bibnamefont
  {Guo}}, \bibinfo {author} {\bibfnamefont {J.}~\bibnamefont {Chang}}, \bibinfo
  {author} {\bibfnamefont {X.}~\bibnamefont {Yao}},\ and\ \bibinfo {author}
  {\bibfnamefont {S.}~\bibnamefont {Gr{\"o}blacher}},\ }\bibfield  {title}
  {\bibinfo {title} {Active-feedback quantum control of an integrated
  low-frequency mechanical resonator},\ }\href
  {https://doi.org/10.1038/s41467-023-40442-3} {\bibfield  {journal} {\bibinfo
  {journal} {Nature Communications}\ }\textbf {\bibinfo {volume} {14}},\
  \bibinfo {pages} {4721} (\bibinfo {year} {2023})}\BibitemShut {NoStop}%
\bibitem [{\citenamefont {Eastman}\ \emph {et~al.}(2019)\citenamefont
  {Eastman}, \citenamefont {Szigeti}, \citenamefont {Hope},\ and\ \citenamefont
  {Carvalho}}]{eastmanControllingChaosQuantum2019}%
  \BibitemOpen
  \bibfield  {author} {\bibinfo {author} {\bibfnamefont {J.~K.}\ \bibnamefont
  {Eastman}}, \bibinfo {author} {\bibfnamefont {S.~S.}\ \bibnamefont
  {Szigeti}}, \bibinfo {author} {\bibfnamefont {J.~J.}\ \bibnamefont {Hope}},\
  and\ \bibinfo {author} {\bibfnamefont {A.~R.~R.}\ \bibnamefont {Carvalho}},\
  }\bibfield  {title} {\bibinfo {title} {Controlling chaos in the quantum
  regime using adaptive measurements},\ }\href
  {https://doi.org/10.1103/PhysRevA.99.012111} {\bibfield  {journal} {\bibinfo
  {journal} {Physical Review A}\ }\textbf {\bibinfo {volume} {99}},\ \bibinfo
  {pages} {012111} (\bibinfo {year} {2019})}\BibitemShut {NoStop}%
\bibitem [{\citenamefont {Zhu}(2024)}]{Zhu_Feedback_Cooling_Code_quackle}%
  \BibitemOpen
  \bibfield  {author} {\bibinfo {author} {\bibfnamefont {K.}~\bibnamefont
  {Zhu}},\ }\href
  {https://github.com/kozse/Feedback-Cooling-of-Incoherent-Quantum-Mixtures/tree/main}
  {\bibinfo {title} {{Github Repository: Feedback Cooling Code}}} (\bibinfo
  {year} {2024})\BibitemShut {NoStop}%
\bibitem [{\citenamefont {Li}\ \emph {et~al.}(2012)\citenamefont {Li},
  \citenamefont {Sattar},\ and\ \citenamefont
  {Sun}}]{liDeterministicResamplingUnbiased2012}%
  \BibitemOpen
  \bibfield  {author} {\bibinfo {author} {\bibfnamefont {T.}~\bibnamefont
  {Li}}, \bibinfo {author} {\bibfnamefont {T.~P.}\ \bibnamefont {Sattar}},\
  and\ \bibinfo {author} {\bibfnamefont {S.}~\bibnamefont {Sun}},\ }\bibfield
  {title} {\bibinfo {title} {Deterministic resampling: {{Unbiased}} sampling to
  avoid sample impoverishment in particle filters},\ }\href
  {https://doi.org/10.1016/j.sigpro.2011.12.019} {\bibfield  {journal}
  {\bibinfo  {journal} {Signal Processing}\ }\textbf {\bibinfo {volume} {92}},\
  \bibinfo {pages} {1637} (\bibinfo {year} {2012})}\BibitemShut {NoStop}%
\bibitem [{\citenamefont {Kitagawa}(1996)}]{kitagawaMonteCarloFilter1996}%
  \BibitemOpen
  \bibfield  {author} {\bibinfo {author} {\bibfnamefont {G.}~\bibnamefont
  {Kitagawa}},\ }\bibfield  {title} {\bibinfo {title} {Monte {{Carlo Filter}}
  and {{Smoother}} for {{Non-Gaussian Nonlinear State Space Models}}},\ }\href
  {https://doi.org/10.2307/1390750} {\bibfield  {journal} {\bibinfo  {journal}
  {Journal of Computational and Graphical Statistics}\ }\textbf {\bibinfo
  {volume} {5}},\ \bibinfo {pages} {1} (\bibinfo {year} {1996})},\ \Eprint
  {https://arxiv.org/abs/1390750} {1390750} \BibitemShut {NoStop}%
\end{thebibliography}%
\end{document}